\documentclass[11pt,epsfig]{amsart}
\usepackage{multicol}
\usepackage{float}
\usepackage{color}
\usepackage{amsmath, amsfonts, graphicx, amsthm}
\usepackage[T1]{fontenc}
\usepackage{enumerate}
\usepackage{amssymb, amsmath}
\usepackage{color, url}
\usepackage{url}
\usepackage{hyperref}
\usepackage{graphicx}
\usepackage{subfig}

\newtheorem{theorem}{Theorem}
\newtheorem{proposition}[theorem]{Proposition}
\newtheorem{lemma}[theorem]{Lemma}

\newtheorem{remark}{Remark}

\newcommand \eps{\varepsilon}

\def\R0{\mathcal{R}_0}

\newcommand \Rzero{\mathcal{R}_0}

\definecolor{purple}{RGB}{75,0,151}

\begin{document}

\title{Open-minded imitation can achieve near-optimal vaccination coverage}

\author{Ying Xin\textsuperscript{$\S$}, David Gerberry\textsuperscript{$\dagger$}, Winfried Just\textsuperscript{$\S$,$\ddagger$,*}}
\thanks{\textsuperscript{$\S$}Department of Mathematics, Ohio University, Athens, Ohio, 45701, USA}
\thanks{\textsuperscript{$\dagger$}Department of Mathematics, Xavier University, Cincinnati, Ohio, 45207, USA}
\thanks{\textsuperscript{$\ddagger$}Quantitative Biology Institute, Ohio University, Athens, Ohio, 45701, USA}
\thanks{\textsuperscript{*}Corresponding author.  Email: mathjust@gmail.com}

\date{\today}

\begin{abstract}
Studies of voluntary vaccination decisions by rational individuals predict that the population will reach a Nash equilibrium with vaccination coverage
below the societal optimum. Human decision-making involves mechanisms in addition to rational calculations of self-interest, such as imitation of successful others. Previous research had shown that imitation alone cannot achieve better results. Under realistic choices of the parameters it may lead to equilibrium vaccination coverage even below the Nash equilibrium.   However, these findings rely on the widely accepted  use of Fermi functions for modeling the probabilities of switching to another strategy. We consider here a more general functional form of the switching probabilities.   It is consistent with functions that give best fits for empirical data in a widely cited psychological experiment and involves one additional parameter~$\alpha$. This parameter can be loosely interpreted as a degree of open-mindedness. We found both by means of simulations and analytically that sufficiently high values of~$\alpha$ will drive the equilibrium vaccination coverage arbitrarily close to the societal optimum.
\end{abstract}

\maketitle

\tableofcontents

\newpage

\section{Introduction}\label{sec:Intro}

While typically non-life-threatening for healthy individuals,  seasonal influenza is responsible for
	tens of thousands of deaths~\cite{CentersforDiseaseControlandPreventionCDC:2010tw}
and	tens of billions of dollars of lost earnings~\cite{Molinari:2007en} each year in the United States alone.
Given the capacity of universal vaccination to mitigate these consequences and to protect especially vulnerable populations (i.e.~children, pregnant women and the elderly),
the United States Centers for Disease Control and Prevention  recommends that everyone 6 months of age or older get a flu vaccine every year \cite{cdc1}.
Despite the CDC recommendation, roughly half of the American population still does not get a flu shot each year~\cite{cdc2}.

The paper~\cite{Galvani} reports that this empirically observed pattern is in accord with predictions of a model that
conceptualizes vaccination decisions as strategies in a multi-player game. In these so-called \emph{vaccination games,} rational individuals are assumed to
reach a \emph{Nash equilibrium} at which each of them follows a strategy that minimizes each individual's expected cost given what
the other individuals in the population do. At Nash equilibrium, the vaccination coverage will not be sufficient for herd immunity, and the
overall cost to the population will not be minimized.  This misalignment between what is optimal for a society (vaccination levels at the herd immunity threshold) and what is optimal for individuals (freeloading on the rest of the population's vaccination) is often referred to as the \emph{vaccination dilemma} \cite{Nowak}.   The vaccination dilemma was first described  in~\cite{0th-pioneering}, and then later independently
in~\cite{first-pioneer}, albeit not phrased in game-theoretic terminology. The first papers that cast this observation in terms of Nash equilibria in  vaccination games were~\cite{An-pioneer} and~\cite{BauchEarnPNAS2004}. Since these seminal papers appeared, the study of vaccination games has mushroomed.
A recent survey article~\cite{StatPhysVacc} includes a total of 777 citations, including one book-length treatment~\cite{VaccGameBook}.

While these general results on vaccination games explain how individual vaccination decisions are likely to lead to lower vaccination coverages than
the societally optimal herd immunity threshold, they are based on several assumptions that will not be satisfied in a real population.
People may not have an accurate perception of the costs of vaccination (such as likelihood of side effects) and infection, and they don't always
behave rationally. Moreover, disease transmission in real populations is strongly dependent on the patterns by which people make contacts. On the one hand,
these factors can  exacerbate the dilemma inherent in voluntary vaccination. On the other hand, this creates opportunities for designing public policy that
would eliminate or alleviate the vaccination dilemma by offering appropriate incentives or effective dissemination of useful information.
Therefore most of the current research on vaccination games focuses on understanding the processes by which people might arrive at their decisions to vaccinate or remain unvaccinated,
and how these processes will influence the resulting vaccination coverage.

In this paper we focus on the role of imitation.
Imitation is prevalent in much of everyday decision-making, in particular when the environment
is complex or largely unknown.  It can be a very successful procedure for finding advantageous strategies in social games~\cite{WhyCopy}.
Social scientists and psychologists have long recognized the importance of imitation, and it has recently moved into the focus of
economists~\cite{ImitationTheory}.  An important concern in the study of imitation is that inertia, resistance to
change, may be present in nearly all decision-making
processes~\cite{ResistanceChange}.

The influential paper~\cite{Nowak} presented a model that incorporates imitation into the decision-making process for vaccinations against
flu-like infections.
In this model it is assumed that during a vaccination campaign that precedes the actual outbreak of the disease,
each focal individual~$i$ independently makes a decision based on comparing his or her own costs $C(i)$ in the preceding season with the cost~$C(j)$ of
one other randomly chosen individual~$j$.  Then~$i$ either follows the same strategy as in the previous season, or switches to $j$'s strategy of the
previous season.  The probability of switching is given by a so-called \emph{Fermi function}
\begin{equation}\label{eqn:Fermi-1}
p_{switch}(i\rightarrow j) = \frac{1}{1 + e^{-\beta(C(i) - C(j))}}.
\end{equation}

Thus the strategy of a much better performing player is readily
adopted, whereas it is unlikely, but not impossible, to adopt
the strategies of worse performing players. The parameter $\beta$
incorporates the uncertainties in the strategy adoption, originating
in either the variation of payoffs or in mistakes in the
decision making. In the limit $\beta \rightarrow 0^+$  player $i$ is unable to retrieve any information from player $j$
and switches to the strategy of $j$ by tossing a fair coin~\cite{FermiText}.

The first use of  Fermi functions for determining the probabilities of switching to another strategy is usually attributed in the literature
to~\cite{Blume}.  They  are examples of ``smoothed imitation'' \cite{GeneralFermi}.
Other examples of smoothed imitation would be functions of the form
\begin{equation}\label{eqn:Fermi-affine}
p_{switch}(i\rightarrow j) = \mu + \frac{\nu}{\alpha + e^{- \beta (C(i) - C(j)) + \gamma}}
\end{equation}
for some suitable choice of parameters $\alpha, \mu$, and $\nu$ that gives probabilities.

There is evidence that switching probabilities as in~\eqref{eqn:Fermi-affine} may be more realistic.
In fact, \cite{StrategyUpdating} reports the results of behavioral experiments on imitation in prisoner's dilemma  games. Analysis of observed behaviors and curve fitting  lead to probabilities of switching to the other strategy of the form~\eqref{eqn:Fermi-affine} with
$\alpha = 1, \nu = 1- \mu$ and
 $\mu = 0.28 \pm 0.07, \beta = 0.67 \pm 0.28,
\gamma = -0.11 \pm 0.23$ for the switch from  cooperation to defection.  For the switch from defection to cooperation they found
$\mu = 0.25 \pm 0.01,
\beta = 0.99 \pm 0.23, \gamma = 0.79 \pm 0.14$.

While in ODE-based models such as \cite{Bauch2005} the likelihood of imitation is often assumed to be proportional to the difference in costs, Fermi functions are used in almost all published discrete-time models of vaccination games with imitation. Sometimes
the cost of the focal player is compared with an average cost of several other players~\cite{Imit-Aver, flu-socialImpact, RealisticDecisions, newIncentives},  but the probability of switching still follows the
pattern of~\eqref{eqn:Fermi-1}.  In view of the results of~\cite{StrategyUpdating} the question naturally arises whether the particular form
of smoothing functions given by~\eqref{eqn:Fermi-1} significantly influences the predictions of models based on it; a question which has received surprisingly little attention in the literature thus far.  A notable exception is~\cite{InertiaToSwitching} where two distinct parameter settings
in~\eqref{eqn:Fermi-affine}
that depend on the current strategy of the focal player were assumed.   For this type of setup, it is intuitively clear that
the equilibrium may shift towards the strategy with the more favorable parameters for being imitated.

In this work, we will demonstrate that the choice of the form of the smoothing function itself can significantly alter the model's predictions even if the parameters in~\eqref{eqn:Fermi-affine} do not depend on the current strategy of the focal player.
Most notably, we show that a suitable choice of the smoothing function alone can drive the system to equilibrium vaccination coverages  that are arbitrarily
close to the societal optimum of herd immunity.

There are four  parameters in~\eqref{eqn:Fermi-affine} in addition to the~$\beta$ of~\eqref{eqn:Fermi-1}, but for mathematical convenience,
we will focus on models with switching probabilities of the form
\begin{equation}\label{eqn:Fermi-alpha}
p_{switch}(i\rightarrow j) = \frac{1}{\alpha + e^{-\beta(C(i) - C(j))}}
\end{equation}
that has only one additional parameter~$\alpha \geq 1$.
As will be shown in Subsection~\ref{subsec:GenFermi} below,
this modified switching probability models a situation where individuals only rarely compare their costs with others, but are eager to switch when they do.
Moreover, for every model that uses~\eqref{eqn:Fermi-affine} with~$\gamma < 0$ there exists
a corresponding model with switching probabilities of the form~\eqref{eqn:Fermi-alpha}
that for the same cost and disease transmission parameters predicts the same equilibria and intervals where the vaccination coverage decreases or increases. The additional parameters of~\eqref{eqn:Fermi-affine} may influence the stability of the equilibria and  how fast
vaccination coverages approach them, but they will not influence the positions of the equilibria.

We focus here on the case where the cost of vaccination is low relative to the cost of the disease (see~\eqref{ineq:cv-ci} below), which is the realistic one for flu vaccinations.  We also assume uniform mixing of the population to
eliminate all aspects of the structure of contact networks that may confound the effects of using the modified switching probability~\eqref{eqn:Fermi-alpha}
in place of~\eqref{eqn:Fermi-1}.
Under these assumptions it is reported in~\cite{Nowak} that when $\alpha = 1$,  the equilibrium
coverage is always even lower than the Nash equilibrium, which is already lower than the societal optimum at herd immunity.
In stark contrast, we found that for all sufficiently large values of~$\beta$, as long as we also choose $\alpha$ large enough relative to~$\beta$, the
equilibrium vaccination coverage predicted by our model can be arbitrarily close to the societally optimal value of herd immunity.

\medskip

\section{Our model}

\subsection{The basic structure of our model}

Our model is a difference equation model that predicts the time evolution of flu vaccination coverage from season to season very similarly to the one
of~\cite{Nowak}. Each time step~$n$
represents a year or flu season, and $V_n$ represents the proportion of individuals in the population who decide to get vaccinated in season~$n$.  Decisions on
whether or not to get vaccinated may depend on individual experience in the previous flu season, the current strategy of a host, and on imitation of one
randomly chosen other host.  These
decisions are assumed to be made by all hosts independently and simultaneously prior to any flu outbreak. They collectively determine the vaccination coverage~$V_n$ in
season number~$n$. After individuals make their vaccination decisions, the probability $x_n = x(V_n)$ of infection of any unvaccinated individual in flu season number~$n$ is then calculated based on a standard SIR model.
As in~\cite{Nowak}, we assume here that the vaccine is 100\% effective so that no vaccinated individual will experience infection.

The model is initialized by randomly assigning strategies for the first season.

The above description is written in the language of an agent-based stochastic process, but we did not actually implement and study the model for
finite populations.  Instead, our version assumes a very large population and is a deterministic compartment-level model, based on expected proportions.

\subsection{Some parameters and implementation details}\label{subsec:parameters-details}

We let $c_v > 0$ denote the cost of vaccination, and $c_i > 0$ denote the cost of infection. Costs are treated as fixed positive numbers here that represent average costs.
In our main result Theorem~\ref{thm:1} we will assume
\begin{equation}\label{ineq:cv-ci}
c_i - c_v > c_v > 0,
\end{equation}
which corresponds in the terminology of~\cite{Nowak} and its follow-up papers to an assumption that the \emph{relative cost} of vaccination $c = \frac{c_v}{c_i} < 0.5$.  This seems realistic for infections like seasonal influenza~\cite{Nowak-electronic}. In our simulations we set $c_v = 1$ and
$c_i = 12$, which gives a relative cost of~$c = \frac{1}{12} = 0.0833$ that approximates the upper range of relative costs that were derived
in~\cite{Nowak-electronic} based on data of~\cite{Galvani} for influenza outbreaks in the U.S.

\medskip

 Before the next flu season $n+1$, each player~$i$ updates his or her strategy as follows:
\begin{itemize}
\item First player~$i$ picks a randomly chosen other player~$j$.
\item Then player~$i$ compares his or her own actual cost~$C(i)$ in the current season to the actual cost~$C(j)$ of player~$j$ in the current season.
\item Then player~$i$ switches to player~$j$'s strategy with probability
\begin{equation}\label{eqn:Fermi}
p_{switch}(i\rightarrow j) = \frac{1}{\alpha + e^{-\beta(C(i) - C(j))}}
\end{equation}
and retains the current strategy with probability~$1 - p_{switch}(i\rightarrow j)$.
\end{itemize}

In the second stage of each season, after all individuals in the population have made their vaccination decisions, there will be a flu outbreak. It is assumed
to develop according to a standard ODE-based SIR model with basic reproductive ratio~$\Rzero > 1$ that is kept fixed over all seasons.
The limit $s_\infty$ of the fraction~$s$ of susceptible individuals as $t \rightarrow \infty$ will satisfy (see, for example~\cite{Diekmann}):
\begin{equation}\label{eqn:finsizes-r=1}
	\ln s_\infty - \ln s_0 = \R0 \left(s_\infty - s_0\right).
\end{equation}

In our model, $s_0 = 1 - V_n$, and  $x_n(V_n) = \frac{s_0 - s_\infty}{s_0}$ is then calculated from~\eqref{eqn:finsizes-r=1}.

\section{Preliminary observations}\label{sec:Observations}

\subsection{Costs, Nash equilibria, and the societal optimum}\label{subsec:Costs-Nash}

The expected costs $c_n^V$ and $c_n^U$ for vaccinators and nonvaccinators in season~$n$ will be:
\begin{equation}\label{eqn:costn-restrict}
c_n^V =  c_v \quad \mbox{and} \quad  c_n^U =  c_i  x_n.
\end{equation}

Let
\begin{equation}\label{eqn:V-HIT}
V_{hit} = 1 - \frac{1}{R_0}
\end{equation}
denote the vaccination coverage at the \emph{herd immunity threshold.}
Then $x$ is a strictly decreasing function on the interval~$[0, V_{hit}]$ such that $x(V_{hit}) = 0$.
At Nash equilibrium, we must have $c_n^V = c_n^U$, and it follows that when $c_i x(0) \geq c_v $, there exists a unique Nash equilibrium
with vaccination coverage $V_{Nash} \in [0,V_{hit})$.

\medskip

Now let us consider  a \emph{societally optimal vaccination coverage~$V_{opt}$.}  This  is supposed to minimize the following function
that represents the average cost to the entire population:
\begin{equation}\label{eqn:PC-def}
PC(V) = c_v V + c_i  (1-V) x,
\end{equation}
where for $V \in [0, V_{hit}]$ the function $x = x(V)$ is the unique solution in the interval~$(0,1)$ of the equation
\begin{equation}\label{eqn:x(V)-r=1}
\begin{split}
1- x &= e^{-R x} \ \ \mbox{for} \ \ R = \Rzero (1-V) \ \ \mbox{so that}\\
1- x &= e^{-\Rzero (1-V)x},
\end{split}
\end{equation}
and for $V \in (V_{hit}, 1]$ we have $x(V) = 0$. This equation can be obtained from~\eqref{eqn:finsizes-r=1} by noting that
$s_0 = 1 -V$ and $\frac{s_\infty}{s_0} = 1 -x$.

\begin{theorem}\label{thm:Nash-suboptimal}
Consider an SIR-model with vaccination and $\Rzero > 1$. Then

\smallskip

\noindent
(a) If $c_v \leq c_i x(0)$, then $V_{Nash} < V_{opt}$.

\smallskip

\noindent
(b) When $c_i \geq c_v$, then $PC(V)$ is strictly decreasing on the interval $[0, V_{hit}]$. In particular, $V_{opt} = V_{hit}$.

\smallskip

\noindent
(c) If $c_v > 2c_i$, then $V_{opt} < V_{hit}$.

\smallskip

\noindent
(d) If $c_i < c_v < 2c_i$, then there exists a critical value $V_{crit} \in [0, V_{hit})$
such that the population cost function $PC(V)$ is strictly decreasing on the interval $[V_{crit}, V_{hit}]$.
\end{theorem}

\medskip

Parts~(a) and~(b)  of Theorem~1 are well-known.  Part~(a) is precisely the \emph{vaccination dilemma} that motives interest in studying
vaccination games.  A proof of part~(b) can be found, for example, in~\cite{Nowak-electronic}. Parts~(c) and~(d) are less well-known, if at all.
To make this preprint reasonably self-contained, we will include a complete proof of the entire Theorem~\ref{thm:Nash-suboptimal} in
the first part of the Appendix~\ref{AppendixA}.

Note that part~(d) has a somewhat paradoxical consequence: Suppose $c_i < c_v < 2c_i$ and $\Rzero$ is large enough so that
we still have $c_ix(0) < c_vV_{hit}$.
Then $V_{opt} < V_{hit}$, as the expected societal cost of not vaccinating anybody is lower than the cost of
vaccinating a fraction~$V_{hit}$ of all individuals.  Assume, moreover, that some misguided mandatory vaccination policy is in place and has already achieved a vaccination
coverage that is close enough to~$V_{hit}$.  Then it becomes actually cost-effective to throw good money after bad and increase vaccination efforts
so as to achieve full herd immunity.

\medskip

\subsection{Generalized Fermi functions}\label{subsec:GenFermi}

In Section~\ref{sec:Intro} we mentioned three functional forms for the probability of focal player~$i$ switching to the strategy of player~$j$.
Here we will discuss in more detail the relationship between these forms and the roles of their parameters.  For convenience,
let us repeat their formulas here. A fairly general form of
smoothed imitation is given by
\begin{equation}\label{eqn:Fermi-affine-rep}
q_{switch}(i\rightarrow j) = \mu + \frac{\nu}{\alpha + e^{- \beta (C(i) - C(j)) + \gamma}},
\end{equation}
where the parameters satisfy the inequalities $\alpha, \nu > 0$, $\mu, \beta \geq 0$.

In this paper we focus on the case $\mu = \gamma = 0$ and $\nu = 1$ with $\alpha \geq 1$, so that:
\begin{equation}\label{eqn:Fermi-alpha-rep}
p_{switch}(i\rightarrow j) = \frac{1}{\alpha + e^{-\beta(C(i) - C(j))}}.
\end{equation}

For $\alpha = 1$ we recover the classical Fermi functions
\begin{equation}\label{eqn:Fermi-1-rep}
p_{switch}(i\rightarrow j) = \frac{1}{1 + e^{-\beta(C(i) - C(j))}}.
\end{equation}

Let us first observe that \eqref{eqn:Fermi-affine-rep} and \eqref{eqn:Fermi-alpha-rep} are equivalent in the sense that they exhibit the same directions of
change and therefore produce the same equilibria with rescaled $\alpha$ and the same $\beta$.

\begin{lemma}\label{lem:correspondence}
Consider two models with the same parameters~$\beta, \Rzero$ and with parameters $\alpha'$ and $\alpha$, respectively, that satisfy
$\alpha' = e^\gamma \alpha$ and $\alpha \geq 1$.    Assume that in the first model
the switching probabilities~$q_{switch}(i\rightarrow j)$ are given by~\eqref{eqn:Fermi-affine-rep} for $\alpha'$, while in the second model the switching probabilities
$p_{switch}(i\rightarrow j)$ are given by~\eqref{eqn:Fermi-alpha-rep} for $\alpha$.
Then for any given vaccination coverage~$V_n$, the sign of~$V_{n+1} - V_n$ in the second model will be the same as the sign
of~$V_{n+1} - V_n$ in the first model.
\end{lemma}

\textbf{Proof:}   Let $V_n$ be the vaccination coverage for season $n$, and let $\Delta(n)$ denote the change of vaccination coverage from season~$n$
to season~$n + 1$. Let us consider the second model first, as it is the simpler one.
In this model the switching probabilities are given by~\eqref{eqn:Fermi-alpha-rep}, so that
\begin{equation*}
    \begin{split}
        \Delta_{model 2}(n) &= V_{n+1} - V_n = (1-V_n)V_n sp, \ \mbox{where}\\
        sp &= \frac{(1-x_n)}{\alpha + e^{\beta c_v}}  + \frac{x_n}{\alpha + e^{-\beta(c_i-c_v)}}
         - \frac{(1-x_n)}{\alpha+e^{-\beta c_v}}  - \frac{x_n}{\alpha+e^{\beta(c_i-c_v)}}.
    \end{split}
\end{equation*}

On the other hand, in the first model where the switching probabilities are given by~\eqref{eqn:Fermi-affine-rep} with $\alpha$ replaced by $\alpha' = e^\gamma \alpha$,
\begin{equation*}
    \begin{split}
        \Delta_{model 1}(n) &= V_{n+1} - V_n = (1-V_n)V_nsq,\ \ \mbox{where} \\
        sq &= (1-x_n)\left(\mu+\frac{\nu}{\alpha' + e^{\beta c_v + \gamma}}\right)+
        x_n\left(\mu + \frac{\nu}{\alpha' + e^{-\beta(c_i-c_v) +\gamma}}\right)\\
        &\ \ \ \ - (1-x_n)\left(\mu+\frac{\nu}{\alpha'+e^{-\beta c_v +\gamma}}\right)
         - x_n\left(\mu+\frac{\nu}{\alpha'+e^{\beta(c_i-c_v)+\gamma}}\right)\\
        &= ((1-x_n) + x_n - (1-x_n) - x_n)\mu + \nu\left(\frac{1}{e^{\gamma}}\right) \, sp\\
        &= \nu\left(\frac{1}{e^{\gamma}}\right) \, sp.
    \end{split}
\end{equation*}

Since  $\nu > 0$, the sign of $V_{n+1}-V_n$ in the two models is always the same. $\Box$

\bigskip

\begin{figure}[H]
\begin{center}
\begin{tabular}{ccc}
\hspace{-1cm}
\includegraphics[scale=0.45]{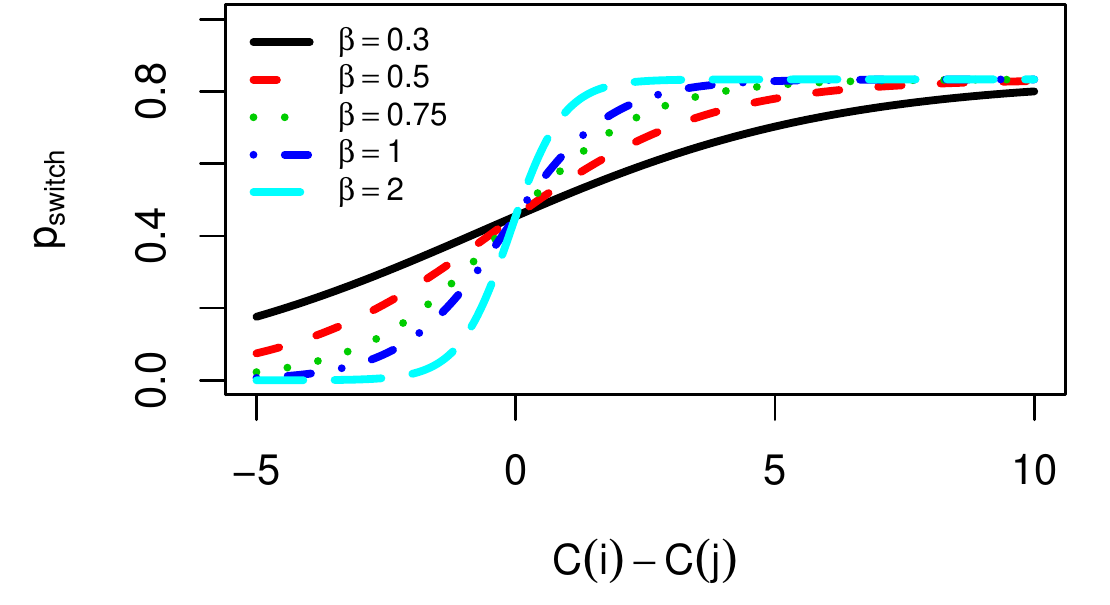}
&
\hspace{0.1cm}
\includegraphics[scale=0.45]{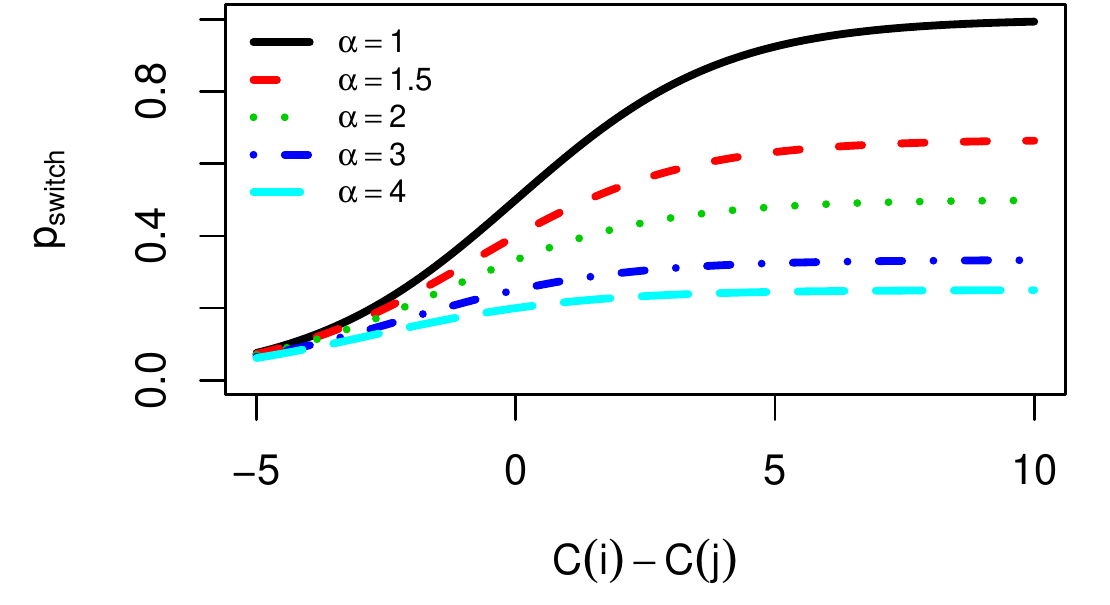}
&
\hspace{0.1cm}
\includegraphics[scale=0.45]{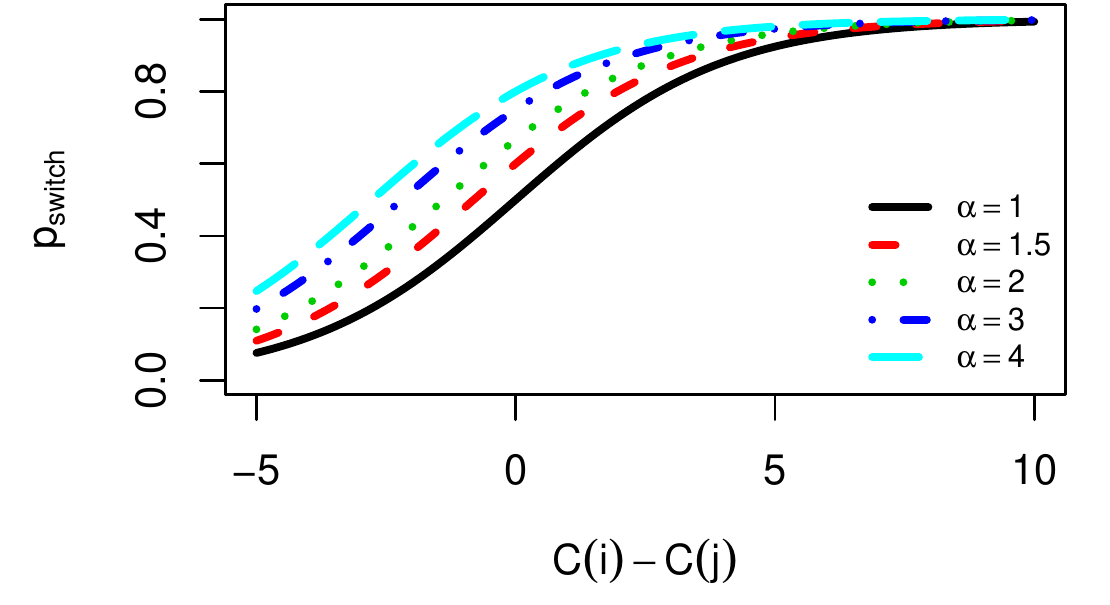}
\end{tabular}
\caption{Dependence of the switching probabilities on the parameters and on the cost difference.
Left panel:  $p_{switch}$ for $\alpha = 1.2$. Middle panel:  $p_{switch}$ for $\beta = 0.5$.  Right panel: $p_{switch\, | \, imitate}$ for $\beta = 0.5$.}
\label{Fig:Fermi}
\end{center}
\end{figure}

The parameter~$\beta$ plays similar roles in our generalized Fermi functions~\eqref{eqn:Fermi-alpha-rep} as in the classical
version~\eqref{eqn:Fermi-1-rep}. As the left panel of Figure~\ref{Fig:Fermi} shows,
when $\beta$ increases, the function $p_{switch}(i\rightarrow j)$ becomes more
like a binary switch that reacts to the sign of the payoff difference~$C(j) - C(i)$.

The parameter~$\alpha$ has two effects.  The first is that high values of~$\alpha$ will make it less likely for the focal player to switch to the other strategy.
See the middle panel of Figure~\ref{Fig:Fermi} for an illustration. In particular, high values of~$\alpha$ will have a stabilizing effect on the interior equilibrium
of our model; see Lemma~\ref{lem:stability-cond} and  Remark~\ref{rem:stability}(a).  Less frequent imitation could also be achieved by choosing low values of~$\nu$, and in view of Lemma~\ref{lem:correspondence} we already
know that the frequency of imitation alone does not change the location of the interior equilibrium.
The main result of this paper, that imitation with our generalized Fermi functions~\eqref{eqn:Fermi-alpha-rep} for sufficiently large~$\alpha$ can give
equilibria that are arbitrarily close to~$V_{hit}$, can be explained only by the second effect of~$\alpha$, which is a more subtle one.
To illustrate this second
effect, let us think of a two-step process for making the decision to imitate.  In the first step, the focal player would make a decision on whether  to
consider imitating another player (with probability~$p_{imitate} = \alpha^{-1}$) or simply do the same as in the last season (with probability~$1 - \alpha^{-1}$).
In the first case, the focal player~$i$ would then compare payoffs~$C(i)$ and~$C(j)$ for one randomly chosen player~$j$ and switch with
conditional probability
\begin{equation}\label{eqn:Fermi-alphalpha}
p_{switch \, | \, imitate}(i\rightarrow j) = \frac{\alpha}{\alpha + e^{-\beta(C(i) - C(j))}}.
\end{equation}

This two-step-procedure is equivalent to the one-step decision given by~\eqref{eqn:Fermi-alpha-rep}.  The right panel of Figure~\ref{Fig:Fermi}
shows how  the function given by~\eqref{eqn:Fermi-alphalpha} depends on~$\alpha$.   We can see that for large~$\alpha$, the focal player is very likely to switch to the other strategy once a decision to consider
imitating has been made, even if that other strategy might be slightly worse than the focal player's current strategy.  For this reason we believe that
high values of~$\alpha$ can be thought of as representing open-mindedness, understood as a willingness to experiment with new strategies unless there is
strong evidence that they are not working well.

\medskip

Let us also mention that the same effect of open-mindedness can be achieved by considering large negative values for~$\gamma$
in~\eqref{eqn:Fermi-affine-rep}.  To see this, recall Lemma~\ref{lem:correspondence} and note that
\begin{equation*}
\begin{split}
q_{switch}(i\rightarrow j) &= \mu + \frac{\nu}{\alpha + e^{- \beta (C(i) - C(j)) + \gamma}} = \mu + \frac{\nu}{\alpha + e^\gamma e^{- \beta(C(i) - C(j))}} \\
&= \mu + \frac{\nu e^{-\gamma}}{\alpha e^{-\gamma} +  e^{- \beta(C(i) - C(j))}}.
\end{split}
\end{equation*}

In~\cite{InertiaToSwitching}, functional forms for the switching probabilities as in~\eqref{eqn:Fermi-affine-rep} for $\mu = 0$ and $\nu = \alpha = 1$
were considered.  The authors suggested that positive values of~$\gamma$ represent inertia and negative values of~$\gamma$ represent ``eagerness
to switch.'' In the context of~\eqref{eqn:Fermi-affine-rep} the relation between $\gamma$ and inertia is not straightforward as  low values of both~$\mu$ and~$\nu$ also give small overall switching probabilities. However, the above calculations do show a direct correspondence between
high values of~$\alpha$ and negative values of~$\gamma$, with both of them having a plausible interpretation in terms of open-mindedness.

\section{Theoretical analysis of the model}\label{sec:theory}

\subsection{Equilibria in our model}\label{subsec:restricted-Nash}

Let~$J: [0,1] \rightarrow [0,1]$ denote the updating function that maps  $V_n$ to $V_{n+1}$.  This function~$J$ is continuous.

\smallskip

If $V_n = 1$, then all individuals vaccinate, and if $V_n = 0$, then nobody vaccinates.  In either case, no player can
switch strategies, since there is nobody in the population to imitate who would follow another strategy.  So $J(0) = 0$, $J(1) = 1$, and $V^{**} := 0$,
$V^{***} := 1$ are always equilibria.

\smallskip

However, we would be more interested in interior equilibria~$V^* \in (0,1)$ for our model.
Note that in our conceptualization of imitation there is always a positive probability that a player will stick with the current strategy rather than imitate
another one, even if that other strategy gave a vastly lower cost.  Thus $J(V) = 0$ only if $V=0$ and $J(V) = 1$ only if $V=1$, which means that the
interior $(0,1)$ is invariant under~$J$. If both $V^{**}$ and
$V^{***}$ are repelling, then the updating function $J$ maps some interval $[\eps, 1-\eps]$ into itself, and  at least one such interior
equilibrium $V^*$ is guaranteed to exist by Brouwer's fixed point theorem.

\smallskip

The equilibrium $V^{***} = 1$ is always repelling. To see this, consider $V_n \in [V_{hit}, 1)$. For very large population sizes~$N$,
we will observe approximately $V_n(1-V_n)N$ comparisons that may induce a player to switch from vaccinating to not vaccinating, and the same number of comparisons
that may induce a player to switch from not vaccinating  to vaccinating. In all these comparisons, players who  vaccinated will have born a positive cost, while
players who did not vaccinate will not have born any cost.  Thus for any choice of the parameters $\alpha, \beta$ we will observe more switches from vaccinating  to not vaccinating than \emph{vice versa,} and it follows that $V_{n+1} < V_n$.

\smallskip

Thus if~$V^*$ exists, it must be in the interval~$(0, V_{hit})$.
Lemma~\ref{lem:V=0-stability-r=1} below shows that an interior equilibrium exists in our model only if~$V^{**} = 0$ is repelling and gives
precise  conditions on the parameters for when this is the case. Lemma~\ref{lem:equilibrium-unique-r=1} below shows that if an interior
equilibrium~$V^*$ exists, it must be unique.

\smallskip

The following theorem is the main result of this paper. It shows that for suitable choices of~$\alpha$ and~$\beta$ the interior equilibrium $V^*$ will be
arbitrarily close to the societal optimum~$V_{hit}$.

%%%%%%%%%%%%%%%%%%%%%%%%%%%%%%%%%%%%%%%%%%%%%%%%%%%%%
%%%%%%%%%%%%%%%%%%%%%%%%%%%%%%%%%%%%%%%%%%%%%%%%%%%%%

\begin{theorem}\label{thm:1}
Fix any $0 \leq V^- < V_{hit}$ and assume the parameters $\alpha, \beta$ of our model satisfy the inequalities
\begin{equation}\label{ineq:beta-cond-r=1}
    1-e^{-2\beta(c_i-c_v)}
    - \frac{2(1-x(V^-))}{x(V^-)}e^{-\beta(c_i-2c_v)} > 0.
\end{equation}

\begin{equation}\label{ineq:alpha-cond-r=1}
    \alpha > \mbox{max}\{1,\ e^{\beta(c_i-c_v)} + e^{-\beta(c_i-c_v)}-2e^{\beta c_v}-2e^{-\beta c_v}\}.
\end{equation}

Then

\smallskip

\begin{itemize}
\item[(a)] If $0 < V_n < V^-$, then $V_{n+1} > V_n$.

\smallskip

\item[(b)] If an interior equilibrium~$V^*$ exists, then $V^* \geq V^-$.
\end{itemize}

\end{theorem}

\noindent
\textbf{Proof:} Note that it suffices to prove part~(a); part~(b) is then an immediate consequence.

Let $V^-, \alpha, \beta$ be as in the assumption, and let $V_n < V^-$.
We will show that the sign of $\Delta(n) := V_{n+1}-V_n$ is positive.
Let $p_{switch}^u$ and  $p_{switch}^v$ denote the conditional probabilities that a player will switch strategies if that player did not vaccinate or did vaccinate in season~$n$, respectively.  Then $p_{switch}^u = V_np_{switch}(u \rightarrow v)$ and  $p_{switch}^v = (1-V_n)p_{switch}(v \rightarrow u)$, so that
\begin{equation*}
    \begin{split}
        \Delta(n) &= V_{n+1} - V_n \\
        &= [1-V_n]p_{switch}^u - V_np_{switch}^v\\
        &= \left[(1-V_n)V_nx_n\frac{1}{\alpha+ e^{-\beta(c_i-c_v)}} + (1-V_n)V_n(1-x_n)\frac{1}{\alpha+e^{-\beta(0-c_v)}}\right] \\
        &\ \ \ \ -
        \left[V_n(1-V_n)x_n\frac{1}{\alpha+e^{-\beta(c_v-c_i)}} + V_n(1-V_n)(1-x_n)\frac{1}{\alpha + e^{-\beta(c_v-0)}}\right]\\
        &= (1-V_n)V_n\left[\frac{x_n}{\alpha+e^{-\beta(c_i-c_v)}} + \frac{1-x_n}{\alpha+e^{\beta c_v}} - \frac{x_n}{\alpha+e^{-\beta(c_v-c_i)}} - \frac{1-x_n}{\alpha+e^{-\beta c_v}}\right]\\
        &= (1-V_n)V_n\left[\frac{x_n(\alpha+e^{-\beta(c_v-c_i)}) - x_n(\alpha+e^{-\beta(c_i-c_v)})}{(\alpha+e^{-\beta(c_i-c_v)})(\alpha+e^{-\beta(c_v-c_i)})}\right] \\
        &\ \ \ \ + (1-V_n)V_n\left[\frac{(1-x_n)(\alpha+e^{-\beta c_v})-(1-x_n)(\alpha+e^{\beta c_v})}{(\alpha+e^{\beta c_v})(\alpha + e^{-\beta c_v})}\right]\\
        &= (1-V_n)V_n\left[\frac{x_n(e^{\beta(c_i-c_v)}-e^{-\beta(c_i-c_v)})}{(\alpha+e^{-\beta(c_i-c_v)})(\alpha+e^{\beta(c_i-c_v)})} + \frac{(1-x_n)(e^{-\beta c_v}-e^{\beta c_v})}{(\alpha + e^{\beta c_v})(\alpha+e^{-\beta c_v})}\right]\\
        &= \frac{(1-V_n)V_n}{(\alpha+e^{\beta c_v})(\alpha + e^{-\beta c_v})}\left[quot(\alpha,\beta)x_n(e^{\beta(c_i-c_v)}-e^{-\beta(c_i-c_v)}) + (1-x_n)(e^{-\beta c_v} - e^{\beta c_v})\right],
    \end{split}
\end{equation*}
where
\begin{equation*}
    quot(\alpha, \beta) = \frac{(\alpha+e^{\beta c_v})(\alpha+e^{-\beta c_v})}{(\alpha+e^{\beta(c_i-c_v)})(\alpha + e^{-\beta(c_i-c_v)})}.
\end{equation*}
Let
\begin{equation*}
    f(\alpha, \beta) = quot(\alpha,\beta)x_n(e^{\beta(c_i-c_v)}-e^{-\beta(c_i-c_v)}) + (1-x_n)(e^{-\beta c_v} - e^{\beta c_v}).
\end{equation*}
Now to show that $\Delta(n) > 0$, it suffices to show that $f(\alpha, \beta) > 0$.

\smallskip

For $\beta, V_n$ as in the assumptions we have $x_n > x(V^-)$, and hence
\begin{equation*}
\begin{split}
    1-e^{-2\beta(c_i-c_v)}
    - \frac{2(1-x(V^-))}{x(V^-)}e^{-\beta(c_i-2c_v)} &> 0,\\
    1-e^{-2\beta(c_i-c_v)}
    - \frac{2(1-x_n)}{x_n}e^{-\beta(c_i-2c_v)} &> 0,\\
    1-e^{-2\beta(c_i-c_v)} + \frac{2(1-x_n)}{x_n}e^{-\beta c_i}
    - \frac{2(1-x_n)}{x_n}e^{-\beta(c_i-2c_v)} &> 0.
\end{split}
\end{equation*}

Then
\begin{equation*}
    \begin{split}
        &\frac{1}{2}x_n(e^{\beta(c_i-c_v)}-e^{-\beta(c_i-c_v)}) + (1-x_n)(e^{-\beta c_v}-e^{\beta c_v})\\
        &= \frac{x_n}{2}e^{\beta(c_i-c_v)}\left[1-e^{-2\beta(c_i-c_v)} + \frac{2(1-x_n)}{x_n}e^{-\beta c_i}
    - \frac{2(1-x_n)}{x_n}e^{-\beta(c_i-2c_v)}\right] > 0.
    \end{split}
\end{equation*}

Moreover, the following inequalities are all equivalent:
\begin{equation*}
    \begin{split}
        &quot(\alpha, \beta) = \frac{(\alpha+e^{\beta c_v})(\alpha+e^{-\beta c_v})}{(\alpha+e^{\beta(c_i-c_v)})(\alpha + e^{-\beta(c_i-c_v)})} > \frac{1}{2}\\
        &2\alpha^2 + (2e^{\beta c_v} + 2e^{-\beta c_v})\alpha + 2 > \alpha^2 + (e^{\beta(c_i-c_v)}+e^{-\beta(c_i-c_v)})\alpha + 1\\
        &\alpha^2 + (2e^{\beta c_v} + 2e^{-\beta c_v}- e^{\beta(c_i-c_v)} - e^{-\beta(c_i-c_v)} )\alpha + 1 > 0
    \end{split}
\end{equation*}

\noindent
Thus for $\alpha > \mbox{max}\{1,\ e^{\beta(c_i-c_v)} + e^{-\beta(c_i-c_v)}-2e^{\beta c_v}-2e^{-\beta c_v}\}$
we have
\begin{equation*}
    quot(\alpha, \beta) > \frac{1}{2}.
\end{equation*}
Then
\begin{equation*}
    \begin{split}
        f(\alpha, \beta) &= quot(\alpha,\beta)x_n(e^{\beta(c_i-c_v)}-e^{-\beta(c_i-c_v)}) + (1-x_n)(e^{-\beta c_v} - e^{\beta c_v})\\
        &> \frac{1}{2}x_n(e^{\beta(c_i-c_v)}-e^{-\beta(c_i-c_v)}) + (1-x_n)(e^{-\beta c_v} - e^{\beta c_v})\\
        &> 0.
    \end{split}
\end{equation*} $\Box$

\medskip

\begin{lemma}\label{lem:equilibrium-unique-r=1}
There can be at most one equilibrium
$V^*$ in the interval $(0,1)$.
\end{lemma}

\noindent
\textbf{Proof:} In this argument we will treat $x$ as a variable that is a function of~$V$ rather than of~$n$. Recall that this function is strictly decreasing, and hence invertible, on the interval $[0, V_{hit}]$.  Consider the functions:
\begin{equation}\label{eqn:gG}
\begin{split}
g(x) &:= quot(\alpha,\beta)(e^{\beta(c_i-c_v)}-e^{-\beta(c_i-c_v)})x + (1-x)(e^{-\beta c_v} - e^{\beta c_v}),\\
G(V) &:= quot(\alpha,\beta)(e^{\beta(c_i-c_v)}-e^{-\beta(c_i-c_v)})x(V) + (1-x(V))(e^{-\beta c_v} - e^{\beta c_v}),
\end{split}
\end{equation}
where $quot(\alpha, \beta)$ is as in the proof of Theorem~\ref{thm:1}.
It follows from our calculations of $\Delta(n)$ in that proof that at an interior equilibrium $V^*$ with $x^* := x(V^*)$ we must have
$g(x^*) = G(V^*) = 0$.

Note that $g(x)$ is a linear function with slope
\begin{equation}\label{eqn:slope-g}
m = quot(\alpha,\beta)(e^{\beta(c_i-c_v)}-e^{-\beta(c_i-c_v)}) + e^{\beta c_v} - e^{-\beta c_v} > 0.
\end{equation}

Thus the system can have at most one interior equilibrium. $\Box$

\medskip

\begin{lemma}\label{lem:V=0-stability-r=1}
 The following conditions are equivalent:

\smallskip

\noindent
(a) An interior equilibrium $V^* \in (0, V_{hit})$ exists.

\smallskip

\noindent
(b) The equilibrium $V^{**} = 0$ is repelling.

\smallskip
(c) One of the  following equivalent inequalities holds:
\begin{equation}\label{ineq:V*-exists-condition-r=1}
\begin{split}
&quot(\alpha,\beta)x(0)(e^{\beta(c_i-c_v)}-e^{-\beta(c_i-c_v)}) + (1-x(0))(e^{-\beta c_v} - e^{\beta c_v}) > 0,\\
&x(0)\left(quot(\alpha,\beta)\left(e^{\beta(c_i-c_v)} -e^{-\beta(c_i-c_v)} \right)+  e^{\beta c_v} - e^{-\beta c_v} \right) > e^{\beta c_v} - e^{-\beta c_v},\\
 x(0) &> \frac{1}{quot(\alpha,\beta)\left(\frac{e^{\beta(c_i-c_v)}-e^{-\beta(c_i-c_v)}}{e^{\beta c_v}-e^{-\beta c_v}}\right) +1}.
\end{split}
\end{equation}

\end{lemma}

\noindent
\textbf{Proof:} Let $g(x), G(V)$ be defined as in~\eqref{eqn:gG}. Then the conditions in part~(c) are simply saying that $g(x(0)) = G(0) > 0$.

Here $x(0) = x(V^{**}) $ is the predicted final size of an outbreak with no vaccination whatsoever. Note that the function~$g(x)$ is linear in~$x$ and increasing
by \eqref{eqn:slope-g}, while $x(V)$ is nonincreasing, and strictly decreasing on~$[0, V_{hit}]$. Thus~$G(V)$ will be strictly decreasing on the
interval~$[0, V_{hit}]$.  Since $x(V_{hit}) = 0$ and thus
\begin{equation*}
G(V_{hit}) = e^{-\beta c_v} - e^{\beta c_v} < 0,
\end{equation*}
it follows from the IVT that $V^*$ exists if, and only if, $G(0) > 0$.

\smallskip

The updating in our model can be entirely understood in terms of
the continuous function~$J$ such that $V_{n+1} = J(V_n)$.   Here and in the next subsection we will work with the following formula for this function:
\begin{equation}\label{eqn:J}
\begin{split}
H(V) &:= \frac{(1-V)V}{(\alpha+e^{\beta c_v})(\alpha + e^{-\beta c_v})},\\
J(V) &:= V + H(V)G(V).
\end{split}
\end{equation}

 Thus when $G(0) > 0$, then $J(\eps) > \eps$ for sufficiently small $\eps > 0$ and $V^{**} = 0$ will be repelling. On the other hand, when $G(0) \leq 0$, then we must have
$G(V) < 0$ for all $V \in (0,1]$, the equilibrium~$V^{**}$ will be locally asymptotically stable and globally attracting on~$[0, 1)$, while $V^*$ does not exist. $\Box$

\bigskip

Note that criterion~\eqref{ineq:V*-exists-condition-r=1} is different from the codition $c_v  \geq c_i x(0)$ that makes not vaccinating the rational choice
for all players.  Similarly to Theorem~\ref{thm:1},  this  implies that in our model we can have an equilibrium vaccination coverage that exceeds the Nash equilibrium. For a numerical example, see Subsubsection~\ref{subsubsec:stabilityV**}.

\subsection{Stability of interior equilibria}\label{subsubsec:restricted-r=1:Stability-V*}

Now let us assume that the interior equilibrium~$V^*$ exists and let us investigate its stability, and also whether trajectories that start near
enough~$V^*$ would approach this equilibrium monotonically. We will work with the function~$J$ defined in~\eqref{eqn:J}.

\smallskip

A sufficient condition for stability of~$V^*$ is given by
\begin{equation}\label{eqn:J-eigen-V*-stable}
-1 < \frac{dJ}{dV}(V^*) < 1.
\end{equation}

Similarly, a sufficient condition for monotone approach to~$V^*$ is given by
\begin{equation}\label{eqn:J-eigen-V*-mono}
0 < \frac{dJ}{dV}(V^*) < 1.
\end{equation}

By differentiating $J$ with respect to~$V$ we find that:
\begin{equation}\label{eqn:dJ}
\begin{split}
\frac{dJ}{dV}(V^*)  &= 1 + \frac{dH}{dV}(V^*)G(V^*) + H(V^*)\frac{dG}{dV}(V^*) \\
& =  1 + \frac{dH}{dV}(V^*)(0) + H(V^*)\frac{dG}{dV}(V^*)  =  1 +  H(V^*)\frac{dG}{dV}(V^*),\\
&=  1 + H(V^*) m \frac{dx}{dV}(V^*) \\
&< 1.
\end{split}
\end{equation}

Since $H(V^*) > 0$ for $V^* \in (0,1)$, the last inequality in~\eqref{eqn:dJ} follows from~\eqref{eqn:slope-g} and the fact that $x(V)$ is a strictly decreasing
function on~$[0, V_{hit}]$; no special assumptions on~$\alpha, \beta$ needed so far.

\smallskip

Now consider the
term~$H(V^*) m \frac{dx}{dV}(V^*)$ of the third line of~\eqref{eqn:dJ}. This term is always negative.  For local stability of~$V^*$ we need
\begin{equation}\label{ineq:loc-stab-HmdV}
-2 \leq H(V^*) m \frac{dx}{dV}(V^*),
\end{equation}
and for monotone approach we need that
\begin{equation}\label{ineq:mono-HmdV}
-1 \leq H(V^*) m \frac{dx}{dV}(V^*).
\end{equation}

It remains to investigate bounds on~$\frac{dx}{dV}(V)$. We can argue here as follows: In the absence of vaccination, $x(R_0)$ is a function of~$R_0$ that for $R_0 >1$ takes the value~$x$ that is the unique solution of the first line of~\eqref{eqn:x(V)-r=1}
in the interval~$(0,1)$. (Here and in the remainder of this subsection we use $R$ to indicate a variable and $R_0$ for the fixed parameter of our model.) When we administer perfectly effective vaccine to a fraction~$V$ of the population, we decrease in effect the expected number of secondary infections caused by an index case in the susceptible population by a factor of~$1 - V$, so that the dynamics boil down to an SIR-model with
\begin{equation}\label{eqn:R0-R0V}
R_0^V = R_0(1-V), \qquad \mbox{or, equivalently,} \qquad \frac{R_0}{R_0^V} = \frac{1}{1-V}.
\end{equation}

Note that, in particular, by solving this equation for~$R_0^{V_{hit}} = 1$, we obtain the herd immunity threshold $V_{hit} = 1 - \frac{1}{R_0}$.
From the chain rule we get
\begin{equation}\label{eqn:chain-dV*}
\frac{dx}{dV}(V^*) = \frac{dx}{dR}(x(R_0^{V^*}))\frac{dR_0^V}{dV}(V^*) = -R_0\frac{dx}{dR}(x(R_0^{V^*})).
\end{equation}

We will now use~$x$ as the variable for the final size of the outbreak in the SIR model with parameter~$R = R_0^V$. By implicitly differentiating
the function $x(R)$ that is defined by the first line of~\eqref{eqn:x(V)-r=1} we get
\begin{equation}\label{eqn:implicit-diff}
\begin{split}
-\frac{dx}{dR} &= (- R_0 \frac{dx}{dR} - x)e^{-Rx} = \left(- R \frac{dx}{dR} - x\right)(1 - x),\\
(R(1-x) -1)\frac{dx}{dR} &= x^2 - x,\\
\frac{dx}{dR} &= \frac{x^2 - x}{R(1-x) -1} = \frac{x}{\frac{1}{1-x} - R}.
\end{split}
\end{equation}

We need to keep in mind that in~\eqref{eqn:implicit-diff} the variable $x$ is a function of~$R$.  So our problem will involve working with bounds for the  expression
\begin{equation}\label{eqn:diff-R}
\begin{split}
-R\frac{dx}{dR} &= \frac{R(x - x^2)}{R(1-x) -1} =  \frac{Rx(1- x)}{R(1-x) -1} = \frac{x}{1 - \frac{1}{R(1-x)}}.
\end{split}
\end{equation}
that we get from combining \eqref{eqn:chain-dV*} with~\eqref{eqn:implicit-diff}.

\begin{proposition}\label{prop:Rx-ineq}
For all $x \in (0,1)$ the following inequalities hold:
\begin{equation*}
    -2< \frac{x}{1-\frac{1}{R(1-x)}} < 0
\end{equation*}
\end{proposition}

\noindent
\textbf{Proof:}  By the first line of \eqref{eqn:x(V)-r=1}, we have
\begin{equation*}
    R = -\frac{\ln(1-x)}{x}.
\end{equation*}
Then the inequalities we want to prove can be written as
\begin{equation*}\label{eqn:-20a}
    -2 < \frac{x}{1+\frac{x}{(1-x)\ln(1-x)}} = \frac{(x-x^2)\ln(1-x)}{(1-x)\ln(1-x)+x} < 0,
\end{equation*}
and the result follows from Proposition~\ref{prop:x-ln-ineq} of the first part of the Appendix~\ref{AppendixA}. $\Box$

\bigskip

This gives us  estimates of~\eqref{eqn:diff-R}, but what we really need are bounds on $H(V^*) m \frac{dx}{dV}(V^*)$.
Let
\begin{equation*}
    S(\alpha, \beta) := \frac{e^{\beta(c_i-c_v)}-e^{-\beta(c_i-c_v)}}{(\alpha+e^{\beta(c_i-c_v)})(\alpha+e^{-\beta(c_i-c_v)})} + \frac{e^{\beta c_v}-e^{-\beta c_v}}{(\alpha+e^{\beta c_v})(\alpha+e^{-\beta c_v})}.
\end{equation*}

By the definition of~$quot(\alpha, \beta)$ we get from~\eqref{eqn:slope-g} and~\eqref{eqn:J} that
\begin{equation}\label{eqn:slope-g-H}
\begin{split}
H(V^*)m &= (1-V^*)V^*\left(\frac{e^{\beta(c_i-c_v)}-e^{-\beta(c_i-c_v)}}{(\alpha+e^{\beta(c_i-c_v)})(\alpha + e^{-\beta(c_i-c_v)})} + \frac{e^{\beta c_v} - e^{-\beta c_v}}{(\alpha+e^{\beta c_v})(\alpha+e^{-\beta c_v})}\right),\\
H(V^*)m &= (1-V^*)V^* S(\alpha, \beta).
\end{split}
\end{equation}

Note that the following inequalities always hold for $0 < V^* < 1$:

\begin{equation}\label{eqn:Hm-approx}
0 < S(\alpha, \beta) < \frac{2}{\alpha} \qquad \mbox{and} \qquad
0 < H(V^*)m < \frac{2}{\alpha}(1-V^*)V^*.
\end{equation}

Then
\begin{equation}\label{eqn:formof-HmdV}
\begin{split}
 H(V^*) m \frac{dx}{dV}(V^*) &= H(V^*) m \left(-R_0\frac{dx}{dR}\left(x\left(R_0^{V^*}\right)\right)\right)\\
 &= H(V^*) m \frac{R_0}{R_0^{V^*}}\left(-R_0^{V^*}\frac{dx}{dR}\left(x\left(R_0^{V^*}\right)\right)\right)\\
 &= (1-V^*)V^*S(\alpha, \beta)  \frac{R_0}{R_0^{V^*}}\left(-R_0^{V^*}\frac{dx}{dR}\left(x\left(R_0^{V^*}\right)\right)\right)\\
 &= (1-V^*)V^*S(\alpha, \beta)  \frac{R_0}{R_0^{V^*}}\frac{x(R_0^{V^*})}{1-\frac{1}{R_0^{V^*}(1-x(R_0^{V^*}))}}\\
 &= V^*S(\alpha, \beta) \frac{x(R_0^{V^*})}{1-\frac{1}{R_0^{V^*}(1-x(R_0^{V^*}))}},
 \end{split}
\end{equation}
where the last line follows from~\eqref{eqn:R0-R0V}.

If $V^* < S(\alpha,\beta)^{-1}$ and $0<x(R_0^{V^*})<1$, then by~\eqref{eqn:formof-HmdV} and Proposition~\ref{prop:Rx-ineq} we will have
\begin{equation*}
    H(V^*)m\frac{dx}{dV}(V^*) = V^*  S(\alpha,\beta) \frac{x(R_0^{V^*})}{1-\frac{1}{R_0^{V^*}(1-x(R_0^{V^*}))}} > -2\\
\end{equation*}

Similarly, if $V^* < \frac{1}{2}S(\alpha,\beta)^{-1}$ and $0<x(R_0^{V^*})<1$, we will have
\begin{equation*}
    H(V^*)m\frac{dx}{dV}(V^*) = V^* S(\alpha,\beta)\frac{x(R_0^{V^*})}{1-\frac{1}{R_0^{V^*}(1-x(R_0^{V^*}))}}> -1.
\end{equation*}

By \eqref{eqn:Hm-approx} we have $S(\alpha, \beta) < \frac{2}{\alpha}$.  Thus whenever $\alpha \geq 2$ we will have
$V^* < 1 < S(\alpha,\beta)^{-1}$.
Similarly, as long as $\alpha \geq 4$, we  have $V^* < 1 < 0.5S(\alpha,\beta)^{-1}$.
This proves the following result:

\begin{lemma}\label{lem:stability-cond}
Consider the restricted model with imitations and parameters such that the interior equilibrium
$V^* \in (0,1)$ exists.  Then
\begin{itemize}
\item $\alpha \geq 2$ is a sufficient condition for local stability of $V^*$.
\item $\alpha \geq 4$ is a sufficient condition for monotone approach to $V^*$.
\end{itemize}
\end{lemma}

\begin{remark}\label{rem:stability}
(a) The conditions in Lemma~\ref{lem:stability-cond} are sufficient, but not necessary.  However, as our results reported
in \ref{subsec:stability-equilibria-imit} show, \emph{some} conditions on the parameters are needed even for stability.
Thus we will leave it as an open problem to find more precise conditions for local stability of~$V^*$ and monotone approach to this equilibrium.

\smallskip

\noindent
(b) It is of interest to observe here that when we extend our model to allow for switching probabilities of the
form~\eqref{eqn:Fermi-affine}, then the expression for the updating function~$J$ changes. As long as $\gamma = 0$ it will take the form
\begin{equation}\label{eqn:J-nu}
J(V) = V + \nu H(V)G(V),
\end{equation}
where the symbols $H$ and~$G$ denote the same funtions as in~\eqref{eqn:J}. This follows from the proof of Lemma~\ref{lem:correspondence}.
While the parameter~$\nu$ will not affect existence and location of the interior equilibrium~$V^*$, it will modify the calculation of  the derivative of
$J$ so that in analogy to~\eqref{eqn:dJ} we obtain
\begin{equation}\label{eqn:dJ-nu}
\frac{dJ}{dV}(V^*)  =  1 + \nu H(V^*) m \frac{dx}{dV}(V^*).
\end{equation}
It follows that in versions of our model that allow larger values of~$\nu$, such as a version that would use~\eqref{eqn:Fermi-alphalpha} instead
of~\eqref{eqn:Fermi-alpha}, we might see instability of~$V^*$ and sustained oscillations even when~$\alpha$ is large.
\end{remark}

\medskip

\section{Numerical results}\label{sec:numerical}

All numerical results included in this section were obtained by setting the cost parameters to $c_v = 1$ and $c_i =12$.
Some details about the software that we used in these explorations can be found in Appendix~B~\ref{sec:software}.

\subsection{Stability of equilibria}\label{subsec:stability-equilibria-imit}

\subsubsection{Stability of $V^{**} = 0$}\label{subsubsec:stabilityV**}

Lemma~\ref{lem:V=0-stability-r=1} predicts that a population that starts with $V(0) \in (0,1)$ will evolve towards the equilibrium $V^{**} = 0$ where nobody vaccinates if, and only if
\begin{equation}\label{ineq:V**-stable}
       x(0) \leq \frac{1}{quot(\alpha,\beta)\left(\frac{e^{\beta(c_i-c_v)}-e^{-\beta(c_i-c_v)}}{e^{\beta c_v}-e^{-\beta c_v}}\right) +1}.
\end{equation}

Interestingly enough, this condition is different from the
inequality
\begin{equation}\label{ineq:nobody-vaccinates}
    x(0) \leq \frac{c_v}{c_i}
\end{equation}
that gives the condition under which not vaccinating is the rational choice for the entire population.

\smallskip

\noindent
Suppose $c_i = 12$ and $c_v = 1$. Then~\eqref{ineq:nobody-vaccinates} becomes
\begin{equation}\label{ineq:nobody-vaccinates-example}
    x(0) \leq \frac{1}{12} \approx 0.0833.
\end{equation}

\begin{figure}[H]
\begin{center}

\includegraphics[width=0.8\linewidth]{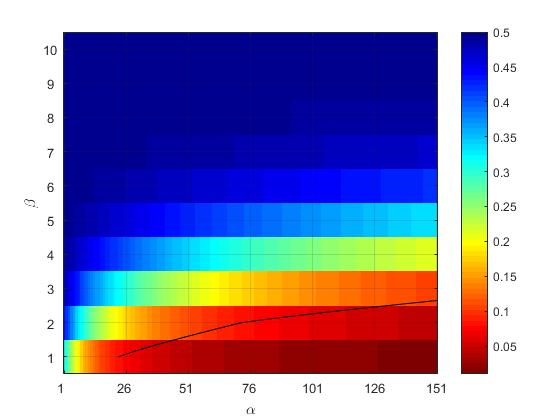}

\caption{Dependence of the right-hand side of~\eqref{ineq:V**-stable}  on $\alpha$ and $\beta$
when $c_i = 12$ and $c_v = 1$. Created with script \texttt{lowerbd.m}.}

\label{Fig:V*6}

\end{center}

\end{figure}

Figure~\ref{Fig:V*6} shows a heat map for the dependence of the right-hand side of~\eqref{ineq:V**-stable}  on $\alpha$ and $\beta$
when $c_i = 12$ and $c_v = 1$. When equality holds in~\eqref{ineq:nobody-vaccinates-example} so that $V_{Nash} = 0$, then
in the region below the black curve the equilibrium~$V^{**} = 0$ becomes repelling, with $V^* \in (0,1)$ predicted to exist.

\medskip

\subsubsection{Counterexamples to stability of~$V^*$}\label{subsubsec:stability-equilibria-imit-counterexamples}
In order to identify regions of the parameter space where $V^*$ is locally asymptotically stable and where approach to this equilibrium will
be monotone, we numerically explored the value of
\begin{equation*}
        H(V^*)m\frac{dx}{dV}(V^*) = V^*S(\alpha,\beta)\frac{x(R_0^{V^*})}{1-\frac{1}{R_0^{V^*}(1-x(R_0^{V^*}))}}
\end{equation*}
with our \textsc{MatLab} scripts \texttt{critical\_R0.m} and \texttt{stability\_and\_approach.m}.

\medskip

\begin{figure}[H]
\begin{center}

\includegraphics[width=0.5\linewidth]{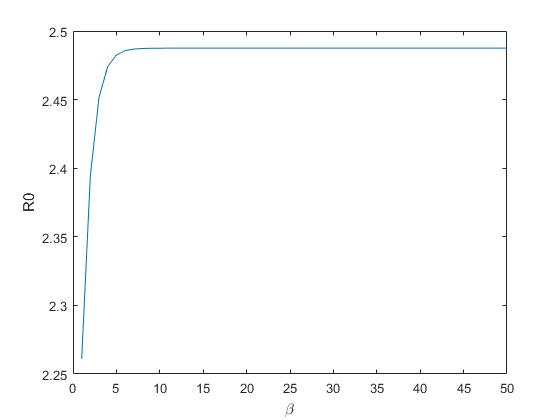}

\caption{Critical value of $R_0$  at which $H(V^*)m\frac{dx}{dV}(V^*) = -1$ for $\alpha = 1$.}

\label{Fig:s1}

\end{center}

\end{figure}

\begin{figure}[H]
\begin{center}

\includegraphics[width=0.5\linewidth]{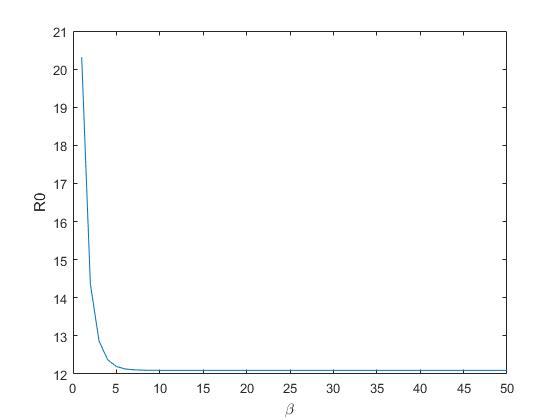}

\caption{Critical value of $R_0$  at which $H(V^*)m\frac{dx}{dV}(V^*) = -1$ for $\alpha = 2$.}

\label{Fig:s2}

\end{center}

\end{figure}

For a range of $\beta$ values, we calculated the critical value of $R_0$  at which $H(V^*)m\frac{dx}{dV}(V^*) = -1$.  For $\alpha =3$, such $R_0$ was not found for $1 \leq \beta \leq 50$. Figures~\ref{Fig:s1} and~\ref{Fig:s2} display the results for $\alpha = 1$ and $\alpha = 2$.

\newpage

\begin{figure}[H]
\begin{center}
\begin{tabular}{ccc}
\hspace{-1cm}
\includegraphics[scale=0.3]{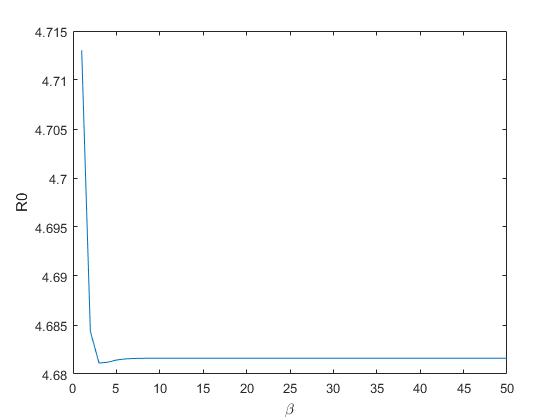}
&
\hspace{0.1cm}
\includegraphics[scale=0.3]{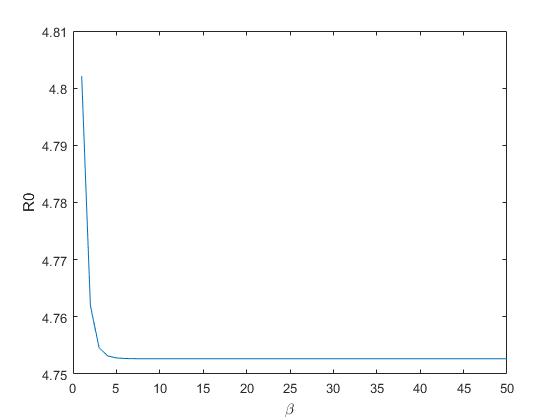}
&
\hspace{0.1cm}
\includegraphics[scale=0.3]{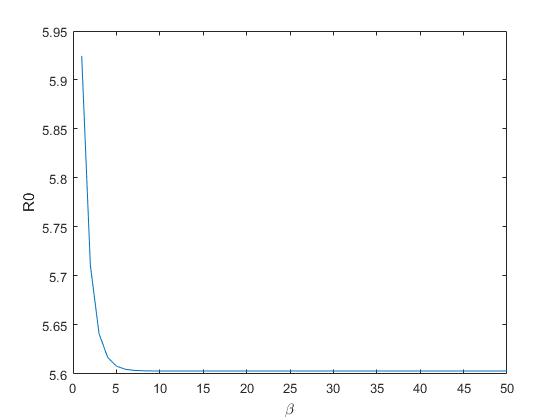}
\end{tabular}
%\caption{}
\end{center}

\begin{center}
\begin{tabular}{ccc}
\hspace{-1cm}
\includegraphics[scale=0.3]{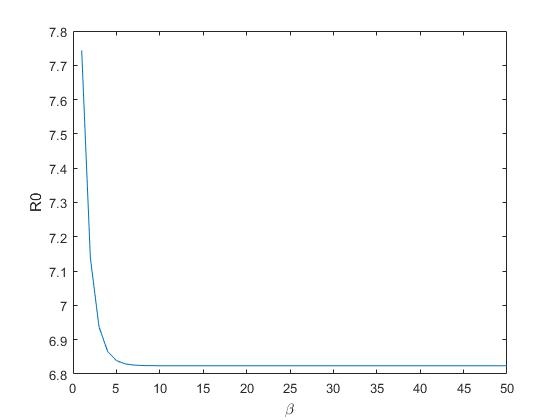}
&
\hspace{0.1cm}
\includegraphics[scale=0.3]{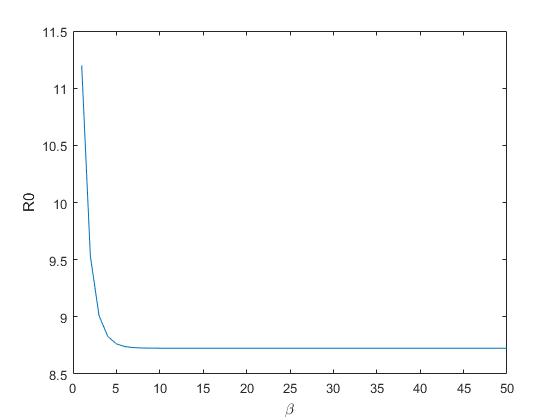}
&
\hspace{0.1cm}
\includegraphics[scale=0.3]{beta_R0_alpha2_neg1.jpg}
\end{tabular}
\caption{Critical value of $R_0$  at which $H(V^*)m\frac{dx}{dV}(V^*) = -1$ for \\
$\alpha = 1, 1.1, 1.2, 1.3, 1.4, 1.5, 1.55, 1.56, 1.565, 1.57, 1.575, 1.58, 1.59, 1.6, 1.7, 1.8, 1.9, 2$\label{Fig:s3.5}}
\end{center}

\end{figure}

In order to see how the seemingly strange inversion of shapes in Figures~\ref{Fig:s1} and~\ref{Fig:s2} occurs, let us consider Figure~\ref{Fig:s3.5} that shows what happens if we gradually increase~$\alpha$.

\begin{figure}[H]
\begin{center}
\begin{tabular}{ccc}
\hspace{-1cm}
\includegraphics[scale=0.3]{beta_R0_alpha1_neg1.jpg}
&
\hspace{0.1cm}
\includegraphics[scale=0.3]{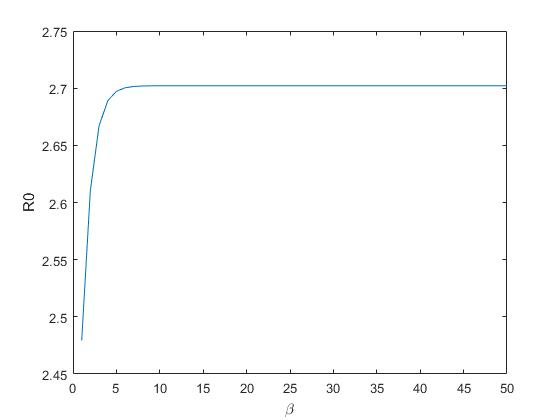}
&
\hspace{0.1cm}
\includegraphics[scale=0.3]{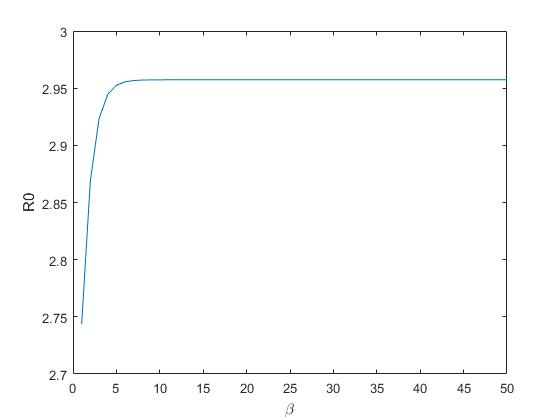}
\end{tabular}
%\caption{}
\end{center}

\begin{center}
\begin{tabular}{ccc}
\hspace{-1cm}
\includegraphics[scale=0.3]{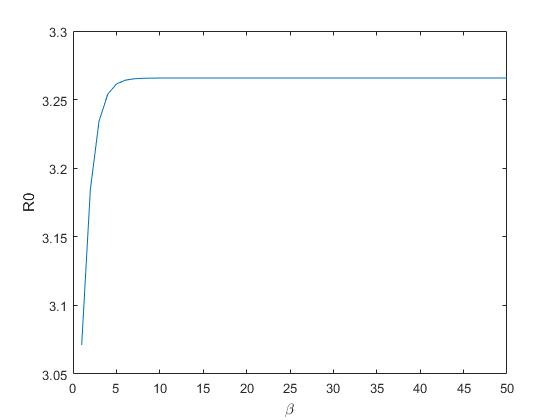}
&
\hspace{0.1cm}
\includegraphics[scale=0.3]{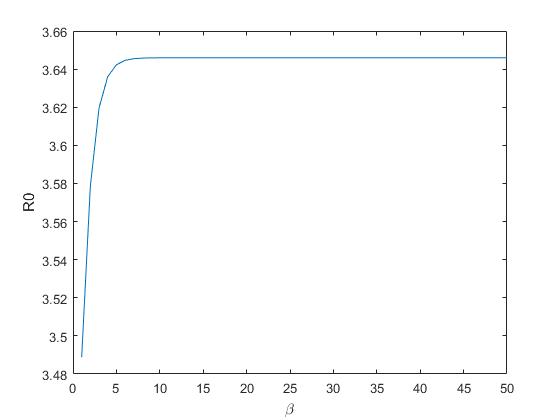}
&
\hspace{0.1cm}
\includegraphics[scale=0.3]{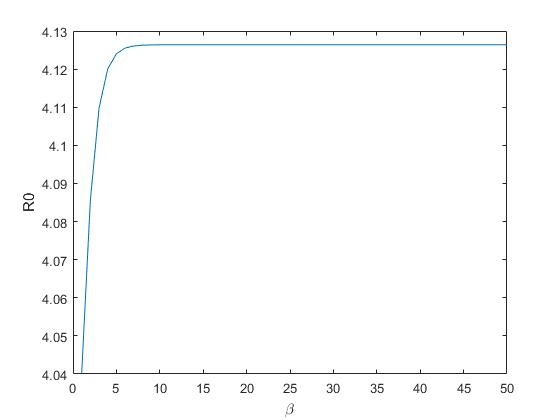}
\end{tabular}
%\caption{}
\end{center}

\begin{center}
\begin{tabular}{ccc}
\hspace{-1cm}
\includegraphics[scale=0.3]{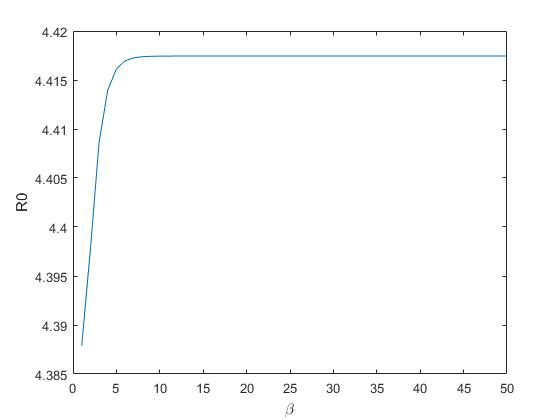}
&
\hspace{0.1cm}
\includegraphics[scale=0.3]{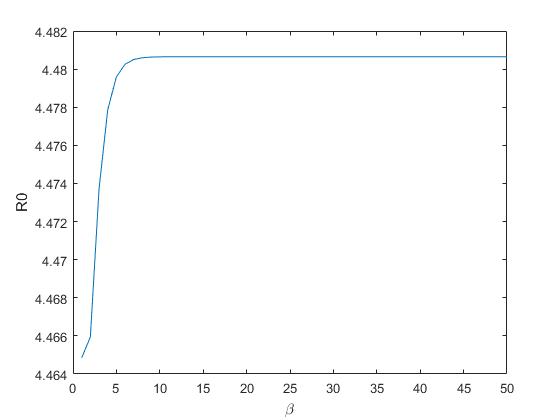}
&
\hspace{0.1cm}
\includegraphics[scale=0.3]{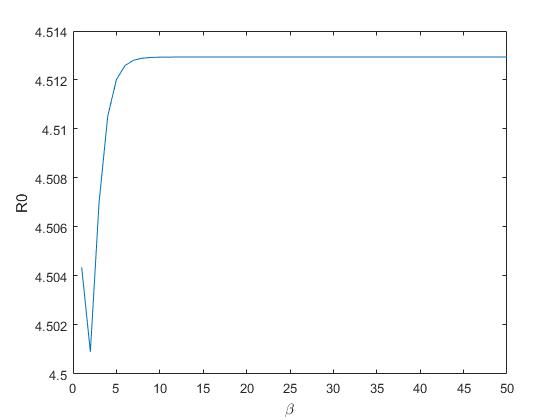}
\end{tabular}
%\caption{}
\end{center}

\begin{center}
\begin{tabular}{ccc}
\hspace{-1cm}
\includegraphics[scale=0.3]{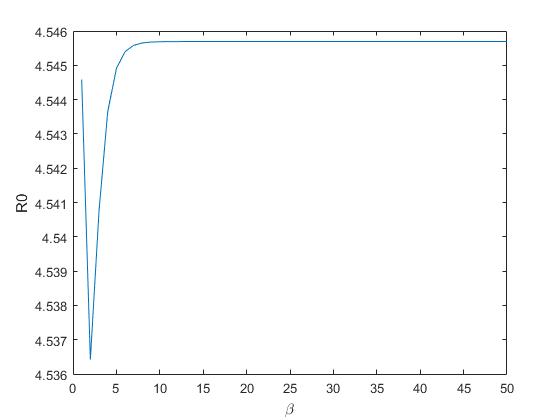}
&
\hspace{0.1cm}
\includegraphics[scale=0.3]{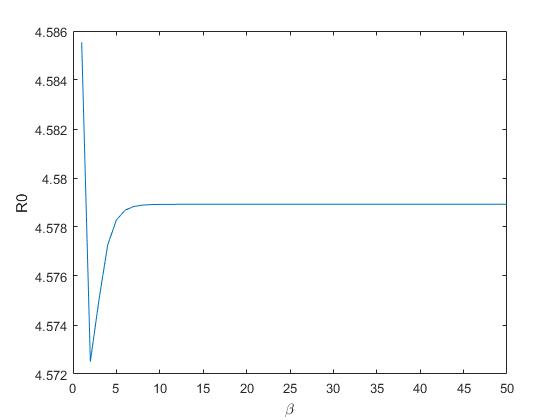}
&
\hspace{0.1cm}
\includegraphics[scale=0.3]{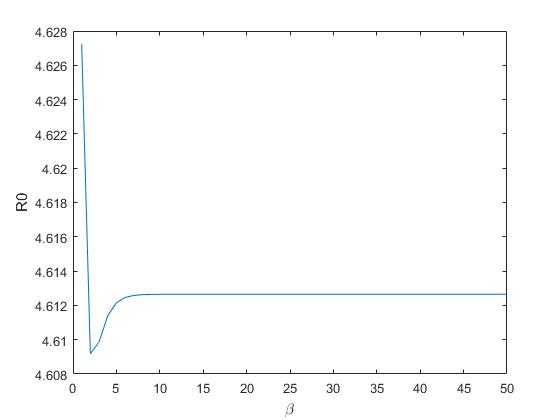}
\end{tabular}
%\caption{}
\end{center}

\end{figure}

\bigskip

Similarly,  for $\alpha = 1$ and for a range of $\beta$ values, we calculated the critical value of $R_0$ at which $H(V^*)m\frac{dx}{dV}(V^*) = -2$.
Figure~\ref{Fig:s3} displays the results:
\begin{figure}[H]
\begin{center}

\includegraphics[width=0.5\linewidth]{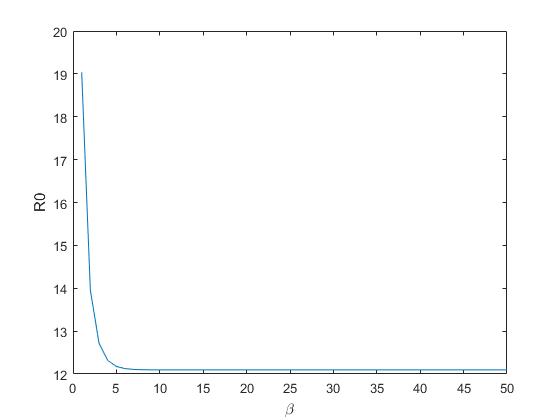}

\caption{Critical value of $R_0$  at which $H(V^*)m\frac{dx}{dV}(V^*) = -2$ for $\alpha = 1$.}

\label{Fig:s3}

\end{center}

\end{figure}

\newpage

These results suggest regions of the parameter space where we would have interior equilibria~$V^*$
that are not approached monotonically and/or where~$V^*$ might be unstable. In order to better visualize these
regions, we color-coded them by distinguishing values in certain relevant intervals.  More specifically, we defined a function
$Int\left(H(V^*)m\frac{dx}{dV}(V^*)\right)$ in the following way:

\begin{itemize}
    \item $Int\left(H(V^*)m\frac{dx}{dV}(V^*)\right) = -4$ if  $H(V^*)m\frac{dx}{dV}(V^*)< -2$,
    \item $Int\left(H(V^*)m\frac{dx}{dV}(V^*)\right) = -1.5$ if $-2 \leq H(V^*)m\frac{dx}{dV}(V^*) < -1.05$,
    \item $Int\left(H(V^*)m\frac{dx}{dV}(V^*)\right) = -0.7$ if $-1.05 \leq H(V^*)m\frac{dx}{dV}(V^*) < -1$,
    \item $Int\left(H(V^*)m\frac{dx}{dV}(V^*)\right) = -0.5$ if  $-1 \leq H(V^*)m\frac{dx}{dV}(V^*) < 0$,
    \item $Int\left(H(V^*)m\frac{dx}{dV}(V^*)\right) = 2$ if \textsc{MatLab} cannot find $V^*$ in $[0,1]$ for given $\alpha$ and $\beta$, or a $V^*$ in $[0,1]$ is found but the corresponding $x(V^*)$ is not in $(0,1)$,
\end{itemize}

\begin{figure}[H]
\begin{center}
\begin{tabular}{cc}
\hspace{-0.5cm}
\includegraphics[scale=0.3]{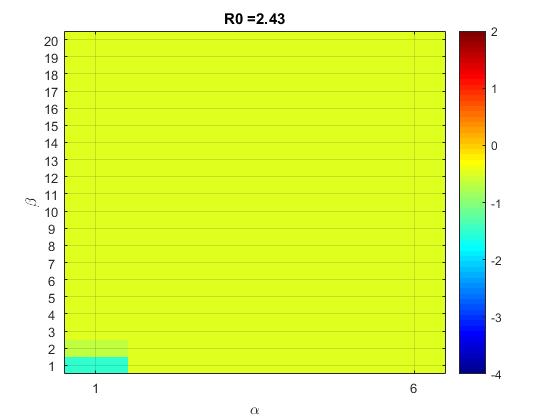}
&
\hspace{0.8cm}
\includegraphics[scale=0.3]{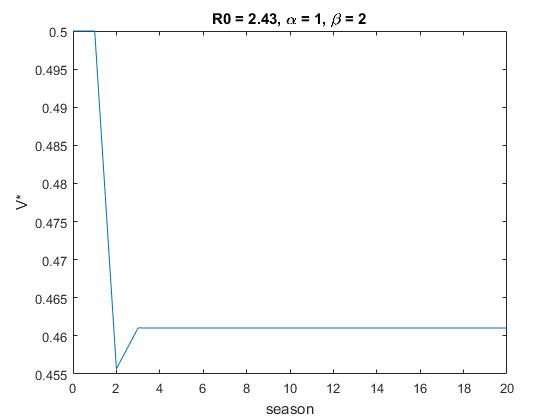}
\end{tabular}
%\caption{}
\end{center}

\begin{center}
\begin{tabular}{cc}
\hspace{-0.5cm}
\includegraphics[scale=0.3]{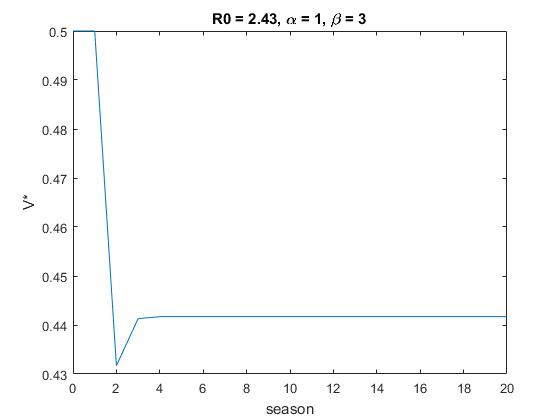}
&
\hspace{0.8cm}
\includegraphics[scale=0.3]{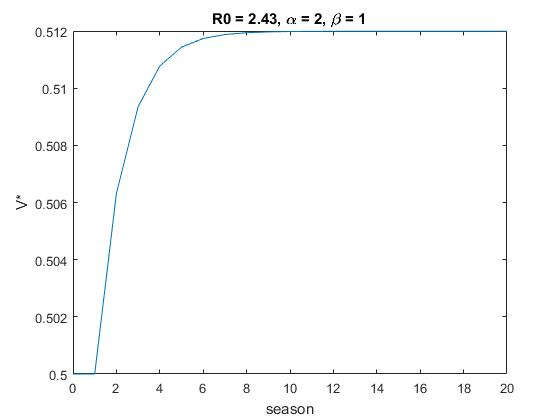}
\end{tabular}
\caption{Color map of $Int\left(H(V^*)m\frac{dx}{dV}(V^*)\right)$ and  sample trajectories for $\Rzero = 2.43$.}
\label{Fig:s4}
\end{center}
\end{figure}

\begin{figure}[H]
\begin{center}
\begin{tabular}{cc}
\hspace{-0.5cm}
\includegraphics[scale=0.3]{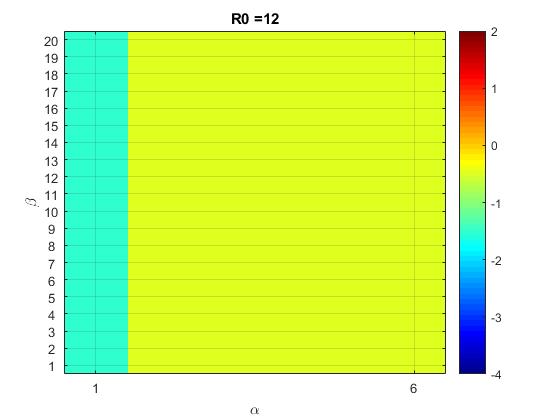}
&
\hspace{0.8cm}
\includegraphics[scale=0.3]{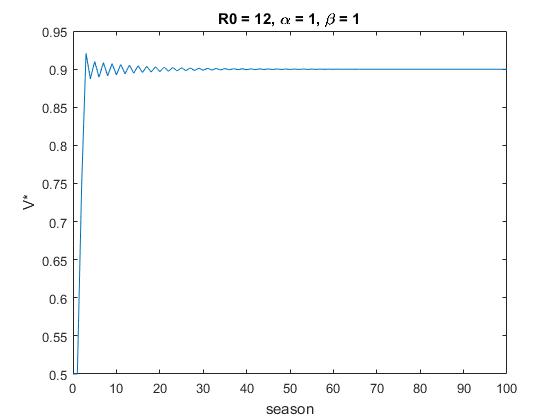}
\end{tabular}
%\caption{}
\end{center}

\begin{center}
\begin{tabular}{cc}
\hspace{-0.5cm}
\includegraphics[scale=0.3]{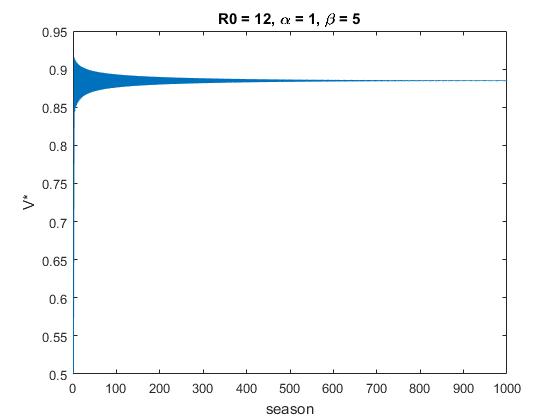}
&
\hspace{0.8cm}
\includegraphics[scale=0.3]{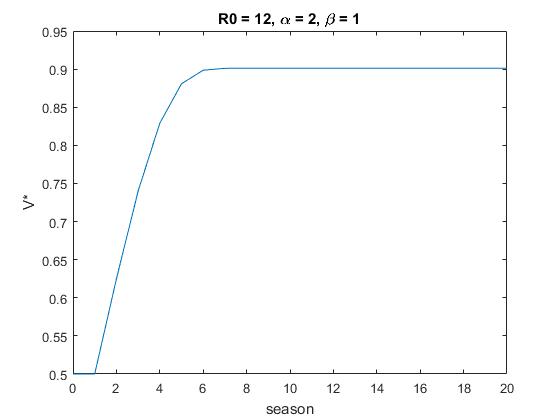}
\end{tabular}
\caption{Color map of $Int\left(H(V^*)m\frac{dx}{dV}(V^*)\right)$ and sample trajectories for $\Rzero = 12$.}
\label{Fig:s5}
\end{center}
\end{figure}

\begin{figure}[H]
\begin{center}
\begin{tabular}{cc}
\hspace{-0.5cm}
\includegraphics[scale=0.3]{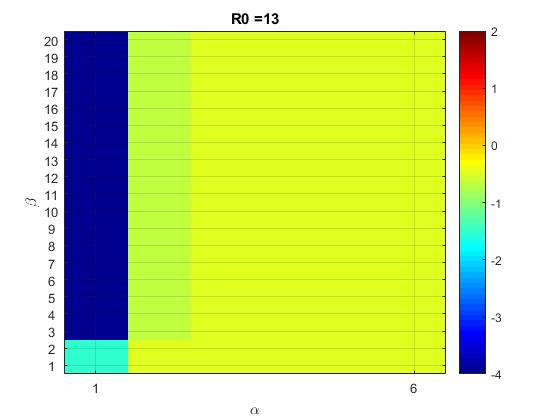}
&
\hspace{0.8cm}
\includegraphics[scale=0.3]{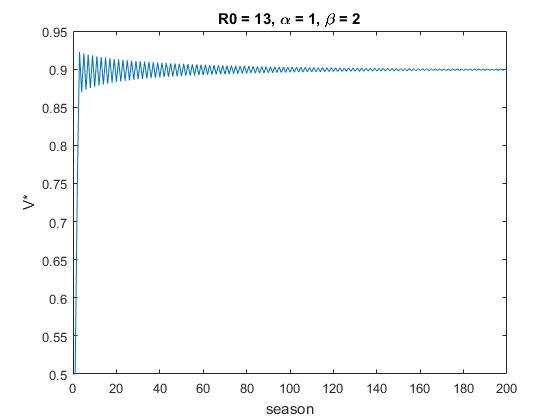}
\end{tabular}
%\caption{}
\end{center}

\begin{center}
\begin{tabular}{cc}
\hspace{-0.5cm}
\includegraphics[scale=0.3]{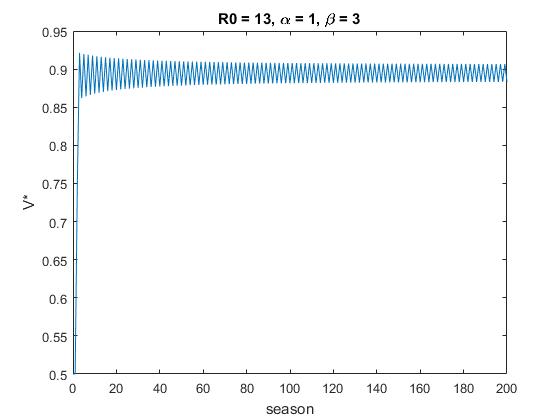}
&
\hspace{0.8cm}
\includegraphics[scale=0.3]{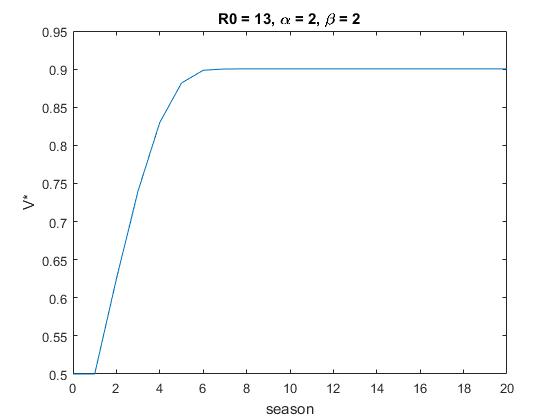}
\end{tabular}
\caption{Color map of $Int\left(H(V^*)m\frac{dx}{dV}(V^*)\right)$ and  sample trajectories for $\Rzero = 13$.}
\label{Fig:s6}
\end{center}
\end{figure}

In Figures~\ref{Fig:s4}--\ref{Fig:s6}, we plot the resulting color-coded partitions of the parameter space, together with some
sample trajectories of $V_n$ in regions of interest that were calculated using our code \texttt{FluVacc}.

\medskip

\subsection{Predicted interior equilibria~$V^*$}\label{subsec:predicted-equilibria-r=1}

The figures in this subsection show the interior equilibria $V^* \in (0,1)$ of the vaccination coverage that we get as outputs
of our script \texttt{V\_G0\_2v\_fsolve.m} that is briefly
described in Subsection~\ref{subsec:restricted-V*-predict-1} of Appendix~B.
Figure~\ref{Fig:V*} shows the dependence of $V^*$ on  $\alpha$ and $\beta$ in the form of heat maps  for selected values of~$\Rzero$.

\begin{figure}[H]
    \centering
    \subfloat[$R_0 = 1.3$]
      {\includegraphics[scale=.45]{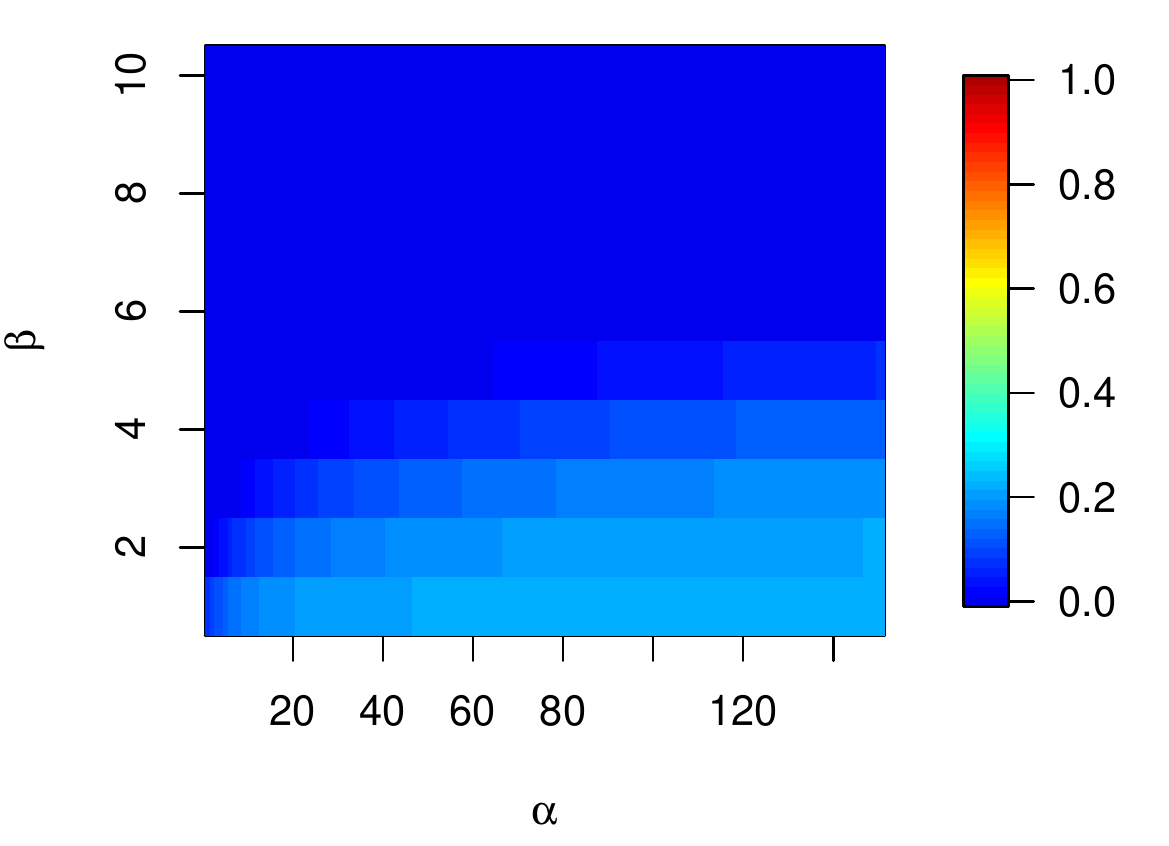}}
   \subfloat[$R_0 = 2.5$]
      {\includegraphics[scale=.45]{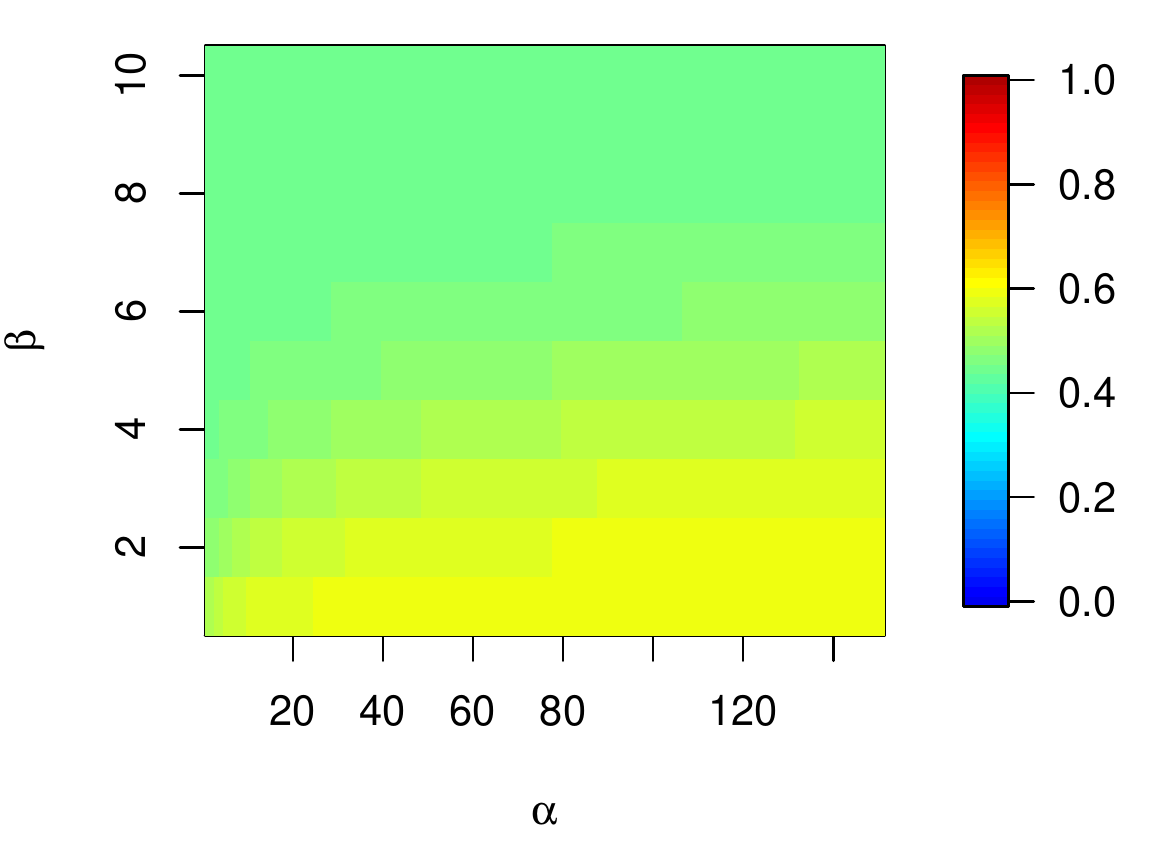}}
 \subfloat[$R_0 = 5.0$]
      {\includegraphics[scale=.45]{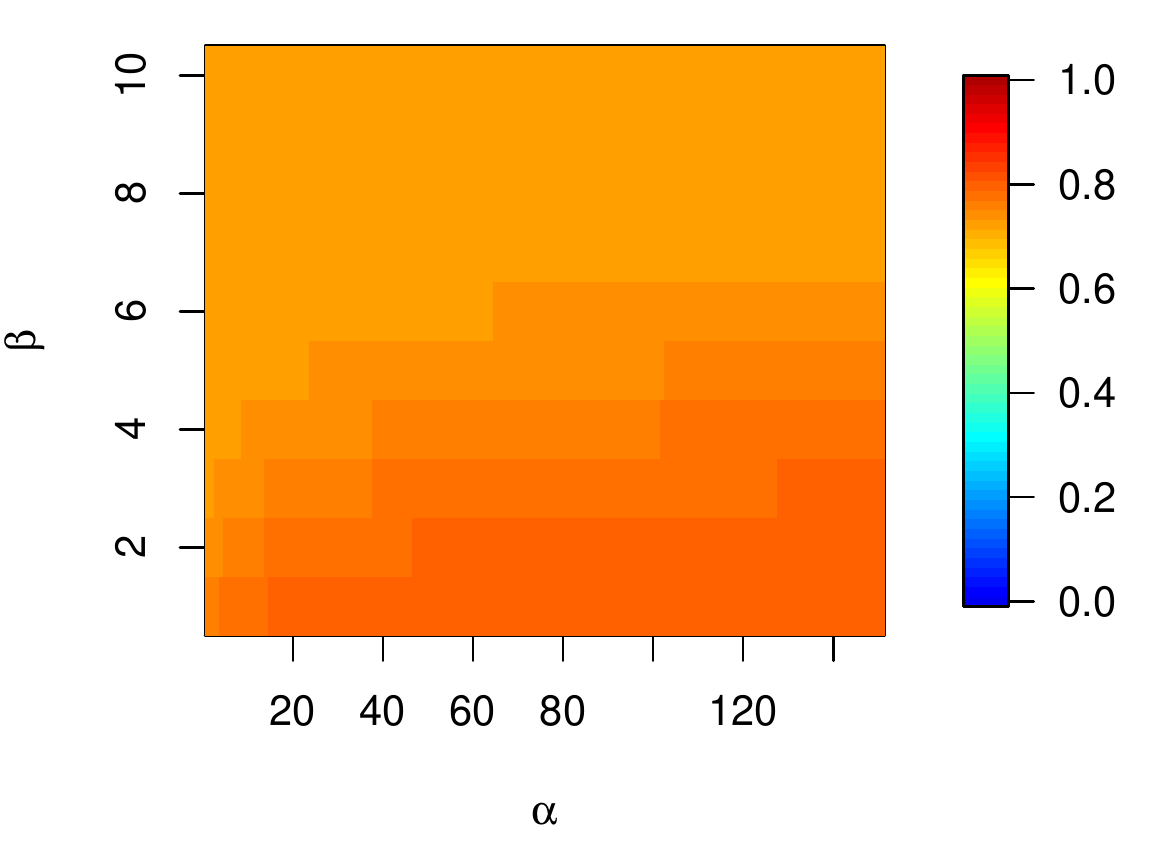}}

    \caption{Dependence of $V^*$ on $\alpha$ and~$\beta$ via theoretical results described in Section~\ref{subsec:restricted-V*-predict-1}
    for $c_v = 1$ and $c_i = 12$.}	
	 \label{Fig:V*}
\end{figure}

\subsection{Simulation results}\label{subsec:simulations-r=1}

\begin{figure}[H]
    \centering
    \subfloat[$R_0 = 1.3$]
      {\includegraphics[scale=.45]{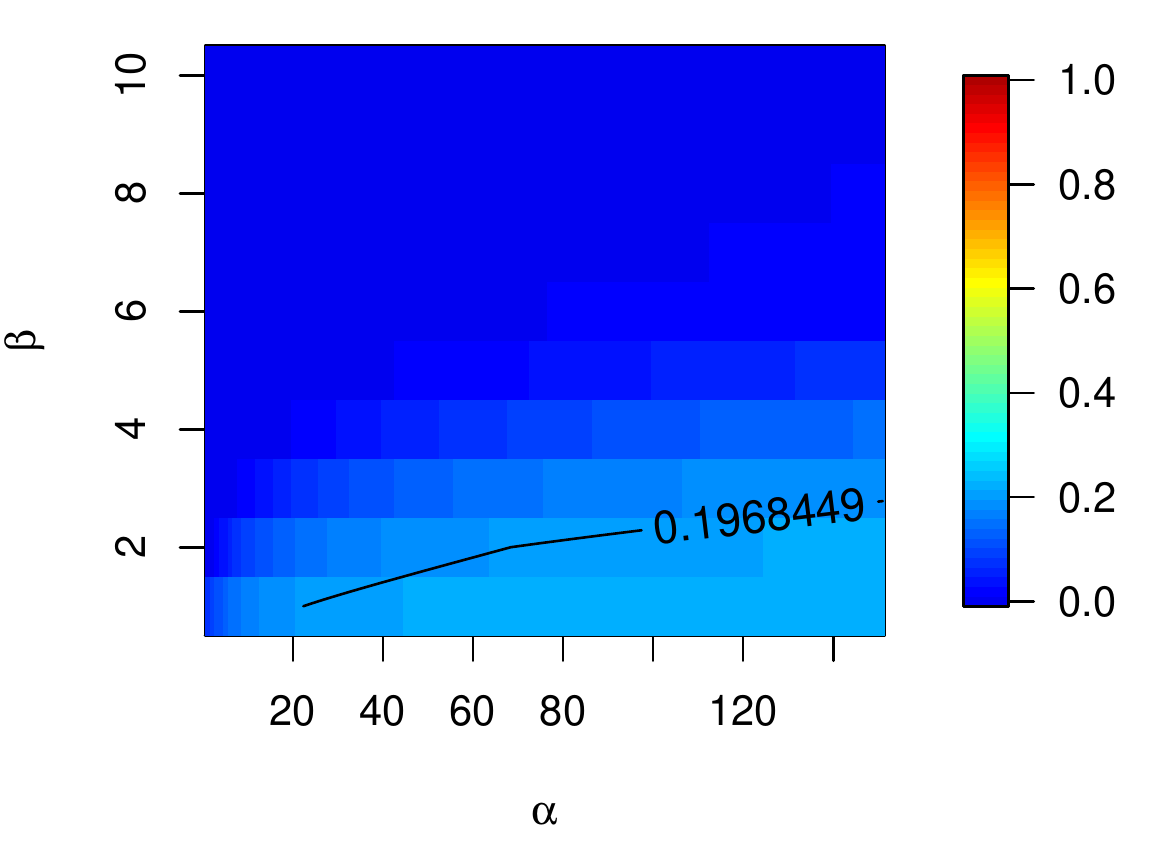}}
   \subfloat[$R_0 = 2.5$]
      {\includegraphics[scale=.45]{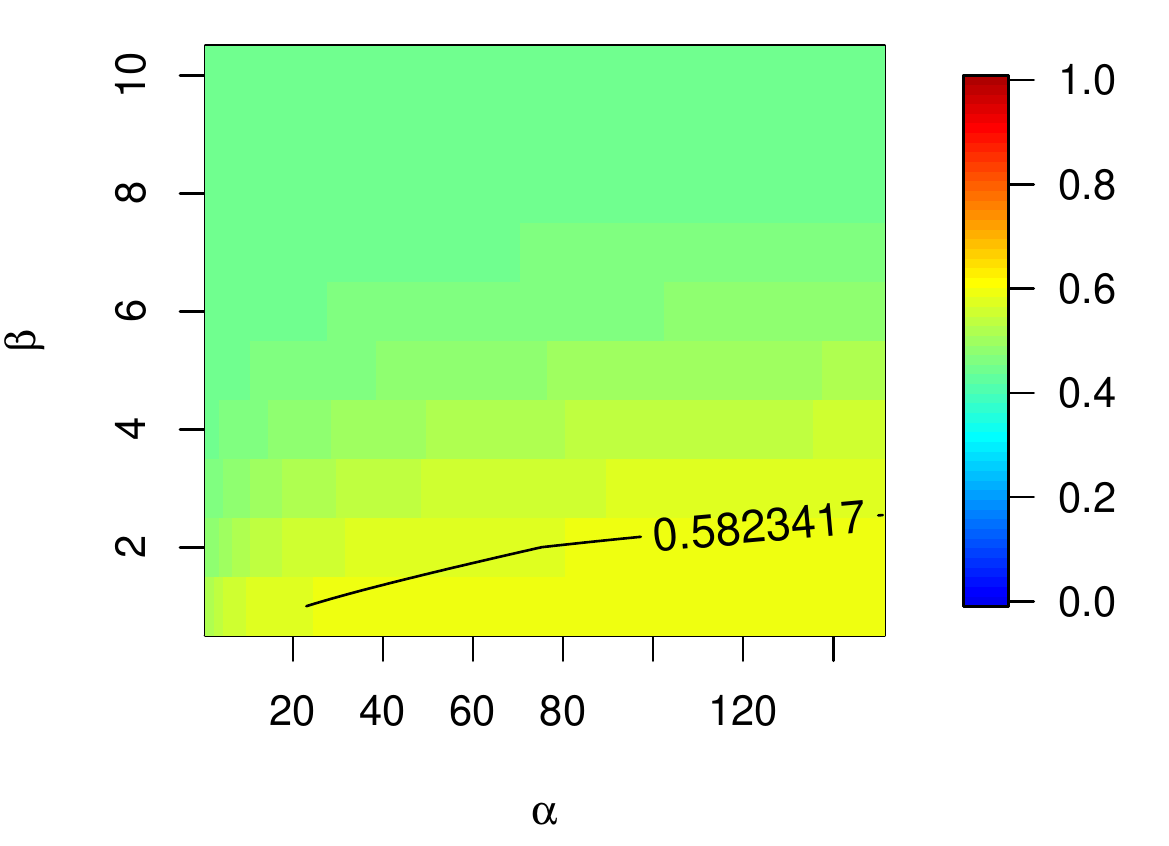}}
 \subfloat[$R_0 = 5.0$]
      {\includegraphics[scale=.45]{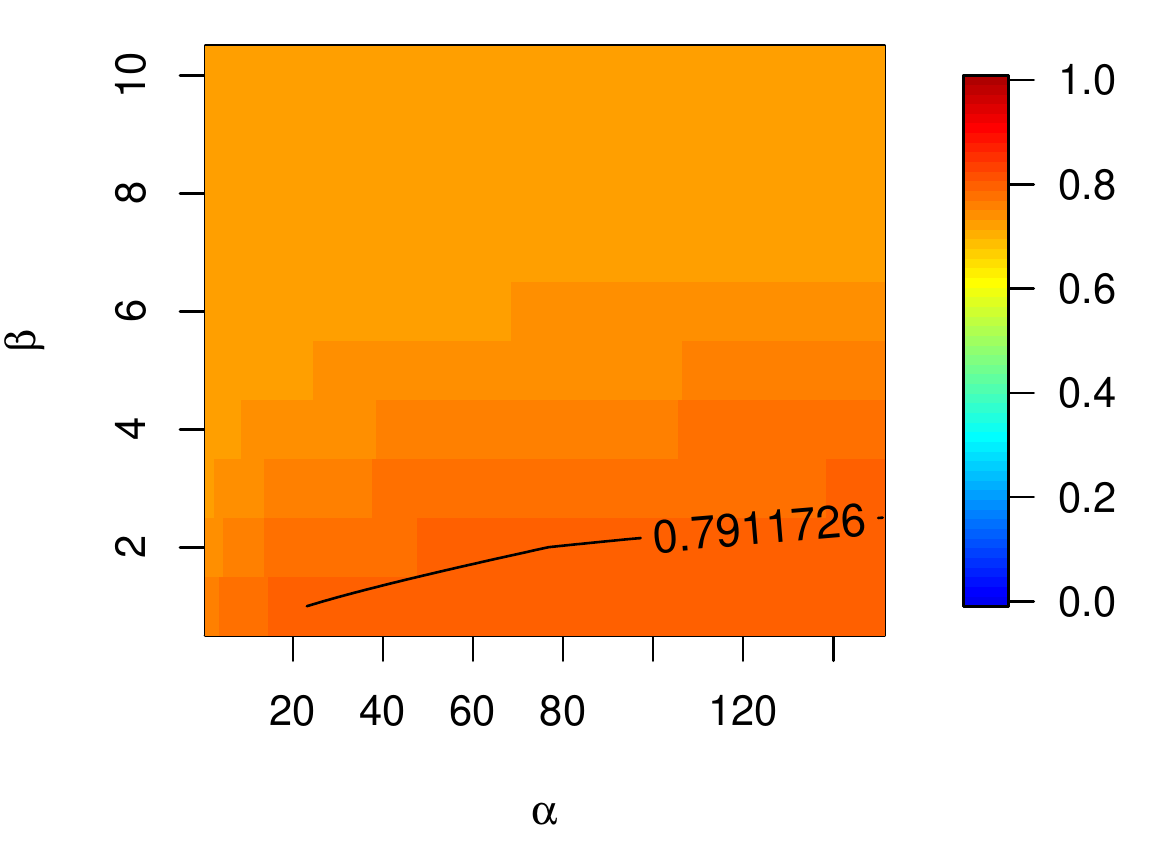}}
    \caption{Dependence of $V^*$ on $\alpha$ and~$\beta$ via numerical simulation of approach to interior equilibrium.
    $c_v = 1$, $c_i = 12$, $V_0 = 0.5$ and $tol = 10^{-5}$.}	
	 \label{Fig:V*sims}
\end{figure}

Figure~\ref{Fig:V*sims} presents the results of direct simulation of our difference equation model described in Section 2 using the script \texttt{simulation\_script.m} that is briefly described in Subsection~\ref{Appendix:batch} of Appendix~B. For each combination of parameters $\alpha$ and $\beta$ the model is simulated to equilibrium (using a stopping criteria of $|V_{n+1} -V_n| < 10^{-5}$) with the resulting vaccination coverage, $V^*$, recorded. Even though this stopping criterion might introduce some small errors, vaccination coverages from direct numerical simulation (Figure~\ref{Fig:V*sims}) match those predicted by the theoretical analysis (Figure~\ref{Fig:V*}) remarkably closely.  The contour on each surface depicts where the model with
imitation matches the vaccine coverage at the Nash equilibrium, which is labeled on the contour.  Therefore, we see that vaccination coverage exceeds that of the Nash equilibrium for combinations of $\alpha$ and $\beta$ that lie below and to the right of this threshold.

\bigskip

\section{Discussion}

The problem of designing effective public policy for inducing people to vaccinate against vaccine-preventable diseases is of great societal
urgency~\cite{VaccGameBook}.  To solve this problem, we need to understand how people really make vaccination decisions and how certain factors that
enter this process influence the outcome.

The work presented here focuses on the role of imitation of successful others, which has been well-documented to be an important component of human
decision-making~\cite{ImitationTheory}.  Our model builds on the version of the model in~\cite{Nowak} that assumed uniform mixing in the population.
The only difference
is including an additional parameter~$\alpha$ in the functional form for the probability of switching to another strategy.
This parameter can be loosely interpreted as a degree of open-mindedness and has parallels
in functional forms of these probabilities that were empirically derived in~\cite{StrategyUpdating}.

Our results demonstrate that for sufficiently
high values of~$\alpha$ the predicted equilibrium coverage will be arbitrarily close to the societally optimal value~$V_{hit}$ that gives herd immunity.
They were confirmed both analytically for large regions of the parameter space in Theorem~\ref{thm:1} and by numerical explorations of the equilibria
predicted in the proof of this theorem (Subsection~\ref{subsec:predicted-equilibria-r=1}) and the equilibria that are being approached in
simulated evolution of the vaccination coverages (Subsection~\ref{subsec:simulations-r=1}).   They
stand in stark contrast to the results reported in~\cite{Nowak} for the standard Fermi functions, which under the uniform
mixing assumption imitation leads to
 vaccination coverage even below the Nash equilibrium when the cost of vaccination is small relative to the cost of infection.  As we deliberately kept our
 model as basic as possible and excluded all other factors, such as community structure, incentives, or misperceptions, these radical differences in the
 predictions can be due only to choosing a high value of~$\alpha$. We conclude that ``open-minded imitation,'' based on switching probabilities
 of the form~\eqref{eqn:Fermi-alpha} with suitable choices of~$\alpha$, provides a possible avenue for attenuating the vaccination dilemma.

\smallskip

The findings presented here open up a number of avenues for future research, in at least four directions. The first would be to try tightening some of our theoretical
results by proving analogues under weaker assumptions.
More specifically, conditions~\eqref{ineq:beta-cond-r=1} and~\eqref{ineq:alpha-cond-r=1}  of Theorem~\ref{thm:1} are sufficient, but not necessary, as
our numerical explorations
in Subsections~\ref{subsec:predicted-equilibria-r=1} and~\ref{subsec:simulations-r=1} indicate that the conclusion of the theorem remains valid under
significantly weaker assumptions.  Similarly,  Lemma~\ref{lem:stability-cond} gives only sufficient conditions for stability of the interior equilibrium~$V^*$
and monotone approach to it.  Our numerical explorations in Subsection~\ref{subsec:stability-equilibria-imit} indicate that these conditions already give us a
qualitatively accurate picture; it would be of some intrinsic mathematical interest to find conditions that are simultaneously sufficient and necessary.

The second direction would be to examine the effect of the parameter~$\alpha$ in extensions of our model that incorporate a number of additional
aspects of voluntary vaccination dynamics that have been deliberately set aside here.  We have already
done some preliminary work on extending our model to the case when the vaccine has an efficacy of less than~100\% and found a similar pattern as the one
reported here in that equilibrium coverage can get arbitrarily close to the societal optimum. However, the dependence of the optimal choice
of~$\alpha$ on~$\beta$ is more complicated than in Theorem~\ref{thm:1} and still needs to be worked out in more detail.
Other aspects that might be incorporated into more detailed versions of our model are restrictions of disease transmission and/or imitation
to edges of contact networks, as has already been studied for the case of classical Fermi functions in~\cite{Nowak} and a number of related papers;
see \cite{AnotherReview} for a review.
Similarly,
one can study the effects of a mixture of rational decision-making and imitation~\cite{public-private, Ndeffo},
incentives~\cite{newIncentives, MoreSubsidies, Subsidies}, misperceptions of
costs~\cite{vaccineScares, Beliefs, Nash-unique, PublicPerceptions, BeyondRational}, altruism
\cite{altruism-Shim, GroupWisdom, other-regard}, peer pressure \cite{flu-socialImpact, Peer},
presence of individuals who remain committed to vaccinating or not vaccinating without ever imitating others~\cite{stubbornVnV, committed},
the effects of other available control measures or treatment
options~\cite{DiseaseInterventions, SI, SPwithNoiseEffects, VaccGamePrevent, newIncentives, Other-Behav-response},
or
variability of~$\Rzero$ from season to season.   One can also study the effects of our parameter~$\alpha$ when imitation is based
on comparison with the average cost of a larger sample of other
individuals~\cite{Imit-Aver, flu-socialImpact, RealisticDecisions, newIncentives}, or when weighted averages of costs over a number of previous seasons are being
compared~\cite{DSwordRational}.

The third direction would be to study applications of our functional form~\eqref{eqn:Fermi-alpha} to domains other than vaccination games.
The paper~\cite{InertiaToSwitching}  considers prisoner's dilemma games in a finite population with switching probabilities of the form
$\frac{1}{1 + e^{- \beta(C(i) - C(j)) - \tau_s}}$, where $\tau_s$ depends on the strategy of the focal player. These can be considered rescaled versions
of~\eqref{eqn:Fermi-alpha} (see the discussion at the end of Subsection~\ref{subsec:GenFermi}).
The authors of~\cite{InertiaToSwitching} obtained a number of analytical results, but their  focus is on fixation probabilities, time to fixation, and stochastic stability of the equilibria, which is different from ours.
 The literature on applications of evolutionary games with the structure of a prisoner's dilemma is vast, and vaccination games are only a small part of it. So
 it seems likely that our version of~\eqref{eqn:Fermi-alpha} or its counterpart in~\cite{InertiaToSwitching} could find many applications outside of
 vaccination games. Let us also remark that there may be some applications to evolutionary computation~\cite{EvoComp} as well. In this field
 the balance between \emph{exploration} and \emph{exploitation} is of paramount importance, and
 our interpretation of high $\alpha$ as open-mindedness bears some resemblance to shifting this balance
 towards the former for high values of this parameter.

Finally, it would be important to get a better understanding of how our parameter~$\alpha$ relates to actual decision-making by real people
and how public policy could enhance more ``open-minded'' decision-making about vaccination in the sense captured by this parameter.
This fourth direction of suggested follow-up work would require a multidisciplinary effort that goes far beyond the realm of mathematical modeling.

\bigskip

\section{Appendix A: Some technical results and proof of Theorem~\ref{thm:Nash-suboptimal}}\label{AppendixA}

\subsection{Some technical results} Here we prove some elementary technical facts that are used in several of our arguments.

\begin{proposition}\label{prop:x-ln-ineq}
For all $x \in (0,1)$ the following inequalities hold:
\begin{equation}\label{eqn:-20}
    -2 <  \frac{(x-x^2)\ln(1-x)}{x+ (1-x)\ln(1-x)} < 0.
\end{equation}
\end{proposition}

\noindent
\textbf{Proof:}  Let $f(x) = (1-x)\ln(1-x)+x$ be the denominator of the above fraction.  Then on $(0,1)$,
\begin{equation*}
    \begin{split}
        f'(x) &= -\ln(1-x) - (1-x)\frac{1}{1-x} + 1 = -\ln(1-x) > 0,\\
        f(x) &> f(0) = 0,\\
        (x-x^2)\ln(1-x) &= x(1-x)\ln(1-x) < 0.
    \end{split}
\end{equation*}
Thus, we have proved the inequality on the right-hand side of~\eqref{eqn:-20}.

\medskip

For the inequality on the left-hand side of \eqref{eqn:-20}, for $x \in (0,1)$,
\begin{equation*}
    \begin{split}
        &\frac{(x-x^2)\ln(1-x)}{(1-x)\ln(1-x)+x} > -2\\
        &\Leftrightarrow \frac{(x-x^2)\ln(1-x)+(2-2x)\ln(1-x)+2x}{f(x)} > 0\\
        &\Leftrightarrow (-x^2-x+2)\ln(1-x) + 2x > 0\\
        &\Leftrightarrow \ln(1-x) > \frac{2x}{x^2+x-2}.
    \end{split}
\end{equation*}
Let $h(x) = g_1(x) - g_2(x)$, where $g_1(x) = \ln(1-x)$ and $g_2(x) = \frac{2x}{x^2+x-2}$. Our goal is to show that $h(x) > 0$ on $(0,1)$.

Note that $g_1(0) = g_2(0) = h(0) = 0$.  On $(0,1)$,
\begin{equation*}
    \begin{split}
        g_1'(x) &= \frac{1}{x-1} = \frac{(x+2)(x^2 + x -2)}{(x^2+x-2)^2} = \frac{x^3 + 3 x^2 - 4}{(x^2+x-2)^2} ,\\
        g_2'(x) &= \frac{-2x^2 - 4}{(x^2+x-2)^2},\\
        h'(x) &= g_1'(x) - g_2'(x) = \frac{x^2(x+5)}{(x^2+x-2)^2} > 0.
    \end{split}
\end{equation*}
We conclude that the left hand side of \eqref{eqn:-20} holds.  $\Box$

\begin{proposition}\label{prop:f-decreases}
The function $f(x) := x + \frac{(x^2-x)\ln(1-x)}{x+(1-x)\ln(1-x)}$ is decreasing on $(0,1)$.
\end{proposition}

\noindent
\textbf{Proof:}
Notice that $f$ can be rewritten as
\begin{equation*}
\begin{split}
    f(x) &= x + \frac{(x^2-x)\ln(1-x)}{x + (1-x)\ln(1-x)} \\
           &= \frac{x^2-(x^2-x)\ln(1-x)+(x^2-x)\ln(1-x)}{x-(x-1)\ln(1-x)} \\
           &= \frac{x^2}{x-(x-1)\ln(1-x)}.
\end{split}
\end{equation*}

To show that $f(x)$ is decreasing on $(0,1)$, we examine its derivative:

\begin{equation*}
\begin{split}
    f'(x) &= \frac{2x(x-(x-1)\ln(1-x)) - x^2(x-(x-1)\ln(1-x))'}{(x-(x-1)\ln(1-x))^2}\\
    &= \frac{2x^2 - (2x^2-2x)\ln(1-x) - x^2(1 - 1 - \ln (1-x)) }{(x-(x-1)\ln(1-x))^2}\\
    &= \frac{2x^2 - (x^2-2x)\ln(1-x)}{(x-(x-1)\ln(1-x))^2}.
    \end{split}
\end{equation*}

The denominator $(x-(x-1)\ln(1-x))^2$ of $f'(x)$ is always positive on $(0,1)$. Thus it suffices to show that the numerator

\begin{equation*}
    g(x) := 2x^2 - (x^2-2x)\ln(1-x)
\end{equation*}
is negative on $(0,1)$.
Observe that
\begin{equation*}
    \begin{split}
        g'(x) &= 4x - \frac{x^2 - 2x}{x-1} - 2(x-1)\ln(1-x),\\
                &= \frac{3x^2-2x}{x-1} - 2(x-1)\ln(1-x),\\
        g''(x) &= \frac{(6x -2)(x-1) -3x^2 + 2x}{(x-1)^2} - 2 - 2\ln(1-x),\\
                 &= \frac{6x^2 -8x +2  -3x^2 + 2x - 2x^2 + 4x -2}{(x-1)^2} - 2\ln(1-x),\\
                 &= \frac{x^2-2x}{(x-1)^2} - 2\ln(1-x) = 1 - \frac{1}{(x-1)^2} - 2\ln (1-x),\\
        g'''(x) &= \frac{2}{(x-1)^3} - \frac{2}{x-1} = \frac{2 - 2x^2 + 4 x -2}{(x-1)^3}\\
                  &= \frac{2x(2-x)}{(x-1)^3} < 0 \ \ \mbox{on}\ (0,1).
    \end{split}
\end{equation*}

Since $g(0) = g'(0) = g''(0) = 0$, we conclude that $g$ is concave down with $g'(x) < 0$ and $g(x) < 0$ on $(0,1)$.  $\Box$

\subsection{Proof of Theorem~\ref{thm:Nash-suboptimal}} We will consider the sign of the derivative
\begin{equation}\label{eqn:PC-prime}
\frac{d\,PC}{d\,V} = c_v  - c_i x + \frac{d\, x}{d\, V}c_i(1-V).
\end{equation}

We already know that under the assumptions of the theorem we must have $V_{Nash} \in (0, V_{hit})$. Then
$c_v  - c_i x= c_v  - c_i x(V_{Nash}) = 0$ by the definition of a Nash equilibrium, and $\frac{d\, x}{d\, V}(V_{Nash}) < 0$. Thus
$\frac{d\,PC}{d\,V}(V_{Nash}) < 0$, which means that the societal cost can be decreased by increasing the vaccination coverage above Nash equilibrium.

For $V \in (0, V_{Nash})$,
\begin{equation*}
    \begin{split}
        \frac{d PC}{d V} &= c_v - c_ix(V) + \frac{d x}{d V}c_i(1-V) \\
        &\leq c_v - c_ix(V_{Nash}) + \frac{d x}{d V}c_i(1-V) = \frac{d x}{d V}c_i(1-V) < 0.
    \end{split}
\end{equation*}
This proves
part~(a).

\medskip

For the proof of parts~(c) and~(d), recall that we already know that $V_{opt} \in [0, V_{hit}]$.  We only need to exclude the possibility that
$V_{opt} = V_{hit}$. It is not immediately clear though whether $\frac{d\, x}{d\, V}(V_{hit})$ exists. Thus we want to investigate
\begin{equation}\label{eqn:PC-prime-Vhit}
\lim_{V \rightarrow V_{hit}^-} \frac{d\,PC}{d\,V}(V) = c_v  +  c_i(1-V)\lim_{V \rightarrow V_{hit}^-}\frac{d\, x}{d\, V}(V).
\end{equation}

By implicitly differentiating the second line of~\eqref{eqn:x(V)-r=1} we obtain:
\begin{equation}\label{eqn:x(V)-prime-r=1}
\begin{split}
\frac{d\,(1- x)}{d\, V} &= \frac{d}{d\, V} e^{-\Rzero (1-V)x}\\
-\frac{d\,x}{d\,V} &= e^{-\Rzero (1-V)x} \frac{d}{d\,V}(-\Rzero (1-V)x)\\
\frac{d\,x}{d\,V} &= \Rzero (1-x)\frac{d}{d\,V}\left(x(1-V)\right)\\
\frac{d\,x}{d\,V} &= \Rzero (1-x)\left((1-V)\frac{d\,x}{d\,V} - x\right)\\
\frac{d\,x}{d\,V} \left(1 - \Rzero (1-x)(1 - V)\right) &= -\Rzero (1-x)x\\
\frac{d\,x}{d\,V} &= \frac{-\Rzero (1-x)x}{\left(1 - \Rzero (1-x)(1 - V)\right)} \\
\frac{d\,x}{d\,V} &= \frac{-\Rzero (1-x)x}{\left(1 + \frac{(1-x)\ln (1-x)}{x}\right)} \\
\frac{d\,x}{d\,V} &= \frac{-\Rzero (x^2-x^3)}{x + (1-x)\ln (1-x))} \\
\lim_{V \rightarrow V_{hit}^-} \frac{d\,x}{d\,V} = \lim_{x \rightarrow 0^+} \frac{d\,x}{d\,V}
&=   \lim_{x \rightarrow 0^+}  \frac{\Rzero (x^3-x^2)}{x + (1-x)\ln (1-x)} \\
&\stackrel{(H)}{=}   \lim_{x \rightarrow 0^+}  \frac{\Rzero (3x^2-2x)}{1 -\ln (1-x) -\frac{1-x}{1-x}} \\
&=    \lim_{x \rightarrow 0^+} \frac{\Rzero (3x-2)x}{-\ln (1-x)}  =   -2\Rzero.
\end{split}
\end{equation}

Now parts~(c) and~(d)  follow from substituting $-2R_0$ for $\lim_{V \rightarrow V_{hit}^-} \frac{d\,x}{d\,V}$ and
$1 - V = 1- V_{hit} =
\frac{1}{\Rzero}$ in~\eqref{eqn:PC-prime-Vhit}.

\medskip

For the proof of part~(b), notice that in view of the above calculations and of the second line of~\eqref{eqn:x(V)-r=1}:
\begin{equation}\label{eqn:PC-prime-new}
\begin{split}
\frac{d\,PC}{d\,V} &= c_v  - c_i \left(x - (1-V)\frac{d\, x}{d\, V}\right)\\
&= c_v  - c_i \left(x - (1-V)\frac{\Rzero (x^3-x^2)}{x + (1-x)\ln (1-x)}\right)\\
&= c_v  - c_i \left(x + \frac{ (x^2-x)\ln (1-x)}{x + (1-x)\ln (1-x)}\right).
\end{split}
\end{equation}

By Proposition~\ref{prop:f-decreases}, the function  $f(x) := x + \frac{ (x^2-x)\ln (1-x)}{x + (1-x)\ln (1-x)}$ is decreasing on $(0,1)$.
Moreover, by  Proposition~\ref{prop:x-ln-ineq},
\begin{equation*}
        0 < \frac{ (x^2-x)\ln (1-x)}{x + (1-x)\ln (1-x)}
\end{equation*} on $(0,1)$.  Thus
\begin{equation*}
        1 < x + \frac{ (x^2-x)\ln (1-x)}{x + (1-x)\ln (1-x)},
\end{equation*}
and it follows that
\begin{equation*}
    \frac{d\,PC}{d\,V} = c_v  - c_i \left(x + \frac{ (x^2-x)\ln (1-x)}{x + (1-x)\ln (1-x)}\right) < c_v - c_i \leq 0
\end{equation*} for $0 \leq V < V_{hit}$ when $c_i \geq c_v$.  $\Box$

\bigskip

\section{Appendix B: Description of our software}\label{sec:software}

We coded several programs that allowed us to numerically explore various aspects of some of our models.  Here we give brief descriptions of how they work.
The codes themselves and complete documentations are available from the authors upon request.

\subsection{Code for predicting stability of interior equilibria}

Here we have two scripts.

\smallskip

Our codes for predicting stability of~$V^*$ is based on estimating the derivative~$\frac{d\, J(V)}{d\, V}$ as caclculated
in~\eqref{eqn:J} and observing that for local stability of~$V^*$ we need
\begin{equation}\label{ineq:loc-stab-HmdV-repeat}
-2 \leq H(V^*) m \frac{dx}{dV}(V^*),
\end{equation}
while for monotone approach to~$V^*$ we need
\begin{equation}\label{ineq:mono-HmdV-repeat}
-1 \leq H(V^*) m \frac{dx}{dV}(V^*).
\end{equation}

\subsubsection{\texttt{critical\_R0.m}}  For given $\alpha$ and a range of $\beta$ values, this script calculates the critical value of $R_0$  at which
\begin{equation*}
    H(V^*)m\frac{dx}{dV}(V^*) = -1 \ \mbox{or}\ -2
\end{equation*}
so that damped oscillations near the interior equilibrium~$V^*$ are predicted to appear/disappear or the local stability of~$V^*$ is predicted to change.
The code plots a graph of the dependence of these critical values on~$\beta$. .

\medskip

\subsubsection{\texttt{stability\_and\_approach.m}}  This script explores stability of equilibria for a user-defined
range of parameter values $(\alpha, \beta)$ when the other parameters are kept fixed.

\smallskip

For given parameters, the script computes the values of $H(V^*)m\frac{dx}{dV}(V^*)$
and compares them with~$-1$ to check for predicted oscillations near~$V^*$
(see~\eqref{ineq:mono-HmdV-repeat})
and with~$-2$ to check for local asymptotic stability of~$V^*$
(see~\eqref{ineq:loc-stab-HmdV-repeat}).

\smallskip

It outputs a color map of  $H(V^*)m\frac{dx}{dV}(V^*)$ by distinguishing values in certain relevant intervals.  More specifically, it displays colors for the
function $Int\left(H(V^*)m\frac{dx}{dV}(V^*)\right)$ that we defined in the following way:

\begin{itemize}
    \item $Int\left(H(V^*)m\frac{dx}{dV}(V^*)\right) = -4$ if  $H(V^*)m\frac{dx}{dV}(V^*)< -2$,
    \item $Int\left(H(V^*)m\frac{dx}{dV}(V^*)\right) = -1.5$ if $-2 \leq H(V^*)m\frac{dx}{dV}(V^*) < -1.05$,
    \item $Int\left(H(V^*)m\frac{dx}{dV}(V^*)\right) = -0.7$ if $-1.05 \leq H(V^*)m\frac{dx}{dV}(V^*) < -1$,
    \item $Int\left(H(V^*)m\frac{dx}{dV}(V^*)\right) = -0.5$ if  $-1 \leq H(V^*)m\frac{dx}{dV}(V^*) < 0$,
    \item $Int\left(H(V^*)m\frac{dx}{dV}(V^*)\right) = 2$ if \textsc{MatLab} cannot find $V^*$ in $[0,1]$ for given $\alpha$ and $\beta$, or a $V^*$ in $[0,1]$ is found but the corresponding $x(V^*)$ is not in $(0,1)$,
\end{itemize}

\medskip

\subsection{Code for numerically predicting equilibria}\label{subsec:restricted-V*-predict}

Here we have two scripts.

\smallskip

\subsubsection{\texttt{lowerbd.m}}  This script calculates the right-hand side of the last line of~\eqref{ineq:V*-exists-condition-r=1} for a user-specified
range of values of the parameters~$\alpha$ and~$\beta$ and graphically displays these values as a heat map. A black curve in the heat map gives
the comparison with $\frac{c_v}{c_i} \approx 0.0833$.

\medskip

\subsubsection{\texttt{V\_G0\_2v\_fsolve.m}}\label{subsec:restricted-V*-predict-1}

Computes the vaccination coverage equilibrium $0 < V^* <1$ that makes $G(V^*) = 0$ if such $V^*$ exists based on the formula for
$\Delta(n) = V_{n+1} - V_n$ that was used in the proof of Theorem~\ref{thm:1}. It performs these calculations for a user-specified range of values
for~$\alpha$ and~$\beta$ while the other model parameters are kept fixed.

\medskip

The script outputs a matrix that contains the predicted values of the interior equilibria~$V^*$ for all pairs~$(\alpha, \beta)$ of parameters explored
and a heatmap that graphically displays these values.

\medskip

For each pair of $\alpha$ and $\beta$ the code uses \texttt{fsolve} to numerically find $V^* \in (0,1)$ so that
 $\Delta(n) = V_{n+1}-V_n = 0$ when $V_n = V^*$.

\medskip

\subsection{Code for simulating  evolution of vaccination coverage}

\subsubsection{\texttt{FluVacc}}
The  package \texttt{FluVacc} of \texttt{MatLab} codes allows the user to simulate the dynamics of our model, as well as in several modifications of it that
are not yet
covered by this preprint. In particular, it allows to monitor time evolution of vaccination coverages and numerically find
Nash equilibria.

\smallskip

Input parameters can be specified by means of a GUI. The simulation results reported in this preprint are based on running a modified version \texttt{FluVacc\_Batch05.m} of the code that allows for
batch processing and uses \texttt{Model} option \texttt{iBf}.
Reported Nash equilibria were found using \texttt{Model} option \texttt{DBf}.

\subsubsection{\texttt{Batch Simulation}}\label{Appendix:batch}

The software package \texttt{Batch Simulation} is primarily made up of scripts from \texttt{FluVacc} modified to allow for batch simulations over ranges of parameters rather than requiring prompted user inputs.
In addition to modified scripts from \texttt{FluVacc} to run the simulations, the package includes a  \textsc{MatLab} script \texttt{simulation\_script.m} to set up the batch of simulations and
the \textsc{R} script \texttt{surfaces.R} that reads in output data of the batch simulations and creates graphics of the results.

\smallskip

\begin{itemize}
\item \texttt{simulation\_script.m}
This is the main file for setting up and running the batch simulations.

\medskip

\item  \texttt{FluVacc\_Batch05.m} Modified form of \texttt{FluVacc.m}.
		Modification include:
		\begin{itemize}
			\item removed GUIs and restricted \texttt{Model} to option \texttt{iBf} and \texttt{OutForm} to option \texttt{Vc}
			\item runs simulation until equilibrium which is defined by vaccination coverage changes by $< tol$ (rather than for a fixed number of seasons) or a maximum number of seasons is reached
			\item has loops to run simulations for ranges of $\alpha$ and $\beta$
			\item saves relevant data to matrix to \texttt{sensitivity\_output}
		\end{itemize}

\item  \texttt{Surfaces.R}  \\ R-script for reading in \texttt{sensitivity\_output} matrix, parsing it into relevant simulation studies and creating graphics of the results.

\smallskip

Takes as input a folder name, reads in simulation result matrix, and make surface plots of relevant simulations in an output folder.
		For each combination of certain input parameters, the scripts creates surface plots in the $\alpha$--$\beta$ plane of
			\begin{itemize}
				\item \texttt{V:} \quad the Vaccination level of the population at fixed point
				\item \texttt{C:} \quad the overall average Cost to the population at fixed point
				\item \texttt{SEA:} \quad the number of SEAsons that each simulation took to reach fixed point
			\end{itemize}	

\end{itemize}

\end{document}